\documentclass[english,11pt]{article}
\usepackage[utf8]{inputenc}
\usepackage{graphicx}
\usepackage{color}
\usepackage{float}
\usepackage{slashed}
\usepackage{amsthm,amsmath,amssymb,mathrsfs}
\usepackage{setspace}
\allowdisplaybreaks
\usepackage{jheppub}
\usepackage{xcolor} 
\usepackage{babel}
\usepackage{booktabs}
\def\thefootnote{\fnsymbol{footnote}}

\def\beq{\begin{equation}}
\def\eeq{\end{equation}}

\begin{document}

\title{The neutrino force in neutrino backgrounds: Spin dependence and parity-violating effects}
\author[a]{Mitrajyoti Ghosh,}
\emailAdd{mghosh2@fsu.edu}
\author[b]{Yuval Grossman,}
\emailAdd{yg73@cornell.edu}
\author[c]{Walter Tangarife,}
\emailAdd{wtangarife@luc.edu}
\author[d]{Xun-Jie Xu,} 
\emailAdd{xuxj@ihep.ac.cn}
\author[b,e]{Bingrong Yu}
\emailAdd{bingrong.yu@cornell.edu}
\affiliation[a]{Department of Physics, Florida State University, Tallahassee, FL 32306-4350, USA}
\affiliation[b]{Department of Physics, LEPP, Cornell University, Ithaca, NY 14853, USA}
\affiliation[c]{Department of Physics, Loyola University Chicago, Chicago, IL 60660, USA} 
\affiliation[d]{Institute of High Energy Physics, Chinese Academy of Sciences, Beijing 100049, China}
\affiliation[e]{Department of Physics, Korea University, Seoul 136-713, Korea}

\abstract{
The neutrino force results from the exchange of a pair of neutrinos. A neutrino
background can significantly influence this force. In this work, we present a comprehensive calculation of the neutrino force in various neutrino backgrounds with spin dependence taken into account. In particular, we calculate
the spin-independent and spin-dependent parity-conserving neutrino forces, in addition to the spin-dependent parity-violating
neutrino forces with and without the presence of a neutrino background for both isotropic and
anisotropic backgrounds. Compared with the vacuum case, the neutrino background can effectively violate Lorentz invariance and lead to additional parity-violating terms that are not suppressed by the velocity of external particles.
We estimate the magnitude of the effect of atomic parity-violation experiments, and it turns out to be well below the current experimental sensitivity.
}

\maketitle
	\def\thefootnote{\arabic{footnote}}
\section{Introduction}	
Neutrinos, with their almost negligible mass, could mediate a long-range force, dubbed the neutrino force, originally conceived in the
	1930s as a possible type of force in nuclei~\cite{papers1930}. Quantitative
	calculations of the neutrino force started in the 1960s, first by Feinberg
	and Sucher~\cite{Feinberg:1968zz},\footnote{At almost the same time, Feynman also considered the neutrino force
		and demonstrated that the three-body neutrino force could resemble
		gravity~\cite{Feynmangravitation}.} followed by a number of subsequent studies addressing various aspects
	of the force~\cite{Feinberg:1989ps,Hsu:1992tg,Grifols:1996fk,Lusignoli:2010gw,LeThien:2019lxh,Segarra:2020rah,Stadnik:2017yge,Ghosh:2019dmi,Bolton:2020xsm,Costantino:2020bei,Xu:2021daf,Dzuba:2022vrv,Munro-Laylim:2022fsv,Ghosh:2022nzo,Blas:2022ovz,VanTilburg:2024tst}.
	For instance, the effect of neutrino masses and mixing was taken
	into account in Refs.~\cite{Grifols:1996fk,Lusignoli:2010gw,LeThien:2019lxh,Segarra:2020rah,Costantino:2020bei};
	the differences between Dirac and Majorana neutrinos were addressed in~\cite{Costantino:2020bei,Segarra:2020rah,Ghosh:2022nzo}; the short-range behavior of the neutrino force was recently investigated in~\cite{Xu:2021daf,Dzuba:2022vrv,Munro-Laylim:2022fsv}; and the cosmological and astrophysical implications have been discussed in~\cite{Fischbach:1996qf,Smirnov:1996vj,Abada:1996nx,Kachelriess:1997cr,Kiers:1997ty,Abada:1998ti,Arafune:1998ft,Orlofsky:2021mmy,Coy:2022cpt}. 

Unlike classical forces such as the Coulomb force, which are generated at the tree level in quantum field theory, the neutrino force is a loop effect. 
It is caused by the exchange of a pair of neutrinos between two test particles. Loop-mediated forces are also called quantum forces.
Generalizations
of the neutrino force with neutrinos being replaced by other light particles (sometimes known as quantum dark forces) have been under active investigation in recent years, see e.g.~\cite{Brax:2017xho,Fichet:2017bng,Costantino:2019ixl,Banks:2020gpu,Brax:2022wrt,Bauer:2023czj,VanTilburg:2024tst,Barbosa:2024tty}.
	
As an inherent feature of quantum forces, the neutrino force can be
influenced by a neutrino background altering the quantum fluctuation of the field~\cite{Ghosh:2022nzo}. Another interesting fact about the neutrino force is that, within the Standard Model (SM), it is the only parity-violating force that could manifest at macroscopic scales~\cite{Ghosh:2019dmi}. Both features have been calculated in our previous work~\cite{Ghosh:2019dmi,Ghosh:2022nzo}
with different focuses. Ref.~\cite{Ghosh:2022nzo} considers only the spin-independent part with a background. Ref.~\cite{Ghosh:2019dmi}
considers the spin-dependent parity-violating force in vacuum. The aim of this work is to perform a complete calculation of the neutrino force in a background, including all spin-dependent terms and background effects. As we will see, in contrast to the vacuum case, the existence of neutrino backgrounds can effectively violate Lorentz invariance and lead to additional sources of parity violation, which are not suppressed by the velocity of external particles.
In addition to the new results obtained in this work, we also notice some discrepancies among existing results in the literature and comment on them. We find that current experimental probes are far from
being capable of detecting the force. Yet, we hope that the comprehensive calculation may be of importance to future long-range force searches.

This paper is structured as follows. In Sec.~\ref{sec:formalism}, we present the general formalism of the neutrino force in an arbitrary neutrino background, taking into account spin-dependent effects.  In Sec.~\ref{sec:isotropic-bkg} and Sec.~\ref{sec:directional-bkg}, we compute the corresponding neutrino forces in specific backgrounds, including the cosmic neutrino background (C$\nu$B), degenerate neutrino gas, and directional and monochromatic neutrino beams.
	In Sec.~\ref{sec:res}, we summarize our results and compare some of them with known results in the literature. Possible experimental probes are discussed in Sec.~\ref{sec:exp}. 
	In Sec.~\ref{sec:conclusion}, we draw conclusions, with some technical details relegated to the appendices.

\section{General formalism}
\label{sec:formalism}
	\begin{figure}[t!]
		\centering
		\includegraphics[scale=0.7]{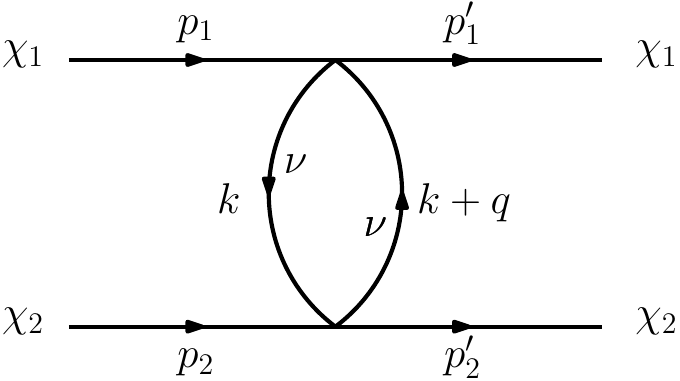}
		\caption{\label{fig:Nuforce}$\chi_1\chi_2\to \chi_1\chi_2$ elastic scattering via the exchange of a pair of neutrinos.}
	\end{figure}
The general four-fermion interaction between neutrinos and some fermion $\chi$ that is weakly charged can be written as 
\begin{eqnarray}
	\mathcal{L}=\frac{G_F}{\sqrt{2}}\left[\bar{\nu}\gamma^\mu\left(1-\gamma_5\right)\nu\right]\left[\bar{\chi}\gamma_\mu\left(g_V^\chi+g_A^\chi\gamma_5\right)\chi\right]\;,
\end{eqnarray}
where $G_F$ is the Fermi constant, $\chi$ can be a lepton, a quark, or any Beyond-the-Standard-Model (BSM) particle, while $g_V^\chi$ and $g_A^\chi$ are the corresponding vector and axial-vector effective couplings to neutrinos. This interaction allows the elastic scattering of two particles $\chi_1$ and $\chi_2$ via exchanging a pair of neutrinos, as shown in Fig.~\ref{fig:Nuforce}. For a very light neutrino mass $m_\nu$, the result is a long-range {\it neutrino} force. For instance, for $m_\nu \sim 0.1~{\rm eV}$, the range of the force is $\ell \sim 1/m_\nu \sim 10^{-4}~{\rm cm}$, which is much larger than the typical range of the weak force. Neglecting the tiny neutrino mass and assuming $g_V^{\chi}=1$,  the spin-independent part of the neutrino force in vacuum is given by
\begin{align}
	\label{eq:2nu}
V_0 (r) = \frac{G_F^2}{4\pi^3 r^5}\;,
\end{align}
where $r$ is the distance between $\chi_1$ and $\chi_2$. Note that the spin-independent part is insensitive to the value of $g_A^{\chi}$. Due to the suppression of $G_F^2$ and the rapidly decreasing scaling of $1/r^5$, $V_0(r)$ is feeble and difficult to probe at large $r$. 

In Ref.~\cite{Ghosh:2022nzo}, we studied the correction of the neutrino force in a general neutrino background. In particular, we found that, in a directional neutrino background with flux $\Phi$ and energy $E_\nu$, there is a large enhancement in the direction parallel to the direction of the neutrino flux,
\begin{align}
	\label{eq:2nu-back}
V_{\rm bkg}(r)\sim \frac{G_F^2 \Phi E_\nu}{r}\;.
\end{align}
Unfortunately, the angular spread of the neutrino flux and test masses substantially smear out the enhancement, making 
it still difficult to probe in practical experiments~\cite{Ghosh:2022nzo,Blas:2022ovz,VanTilburg:2024tst}.
Ref.~\cite{Blas:2022ovz} argued that the finite size of the background neutrino wave packets will significantly reduce the enhancement. We agree with their result. In v2 of Ref.~\cite{Ghosh:2022nzo}, we explicitly showed that the $1/r$ enhancement in Eq.~(\ref{eq:2nu-back}) only exists when  $\alpha^2\lesssim 1/\Delta(E_\nu r)$, where $\alpha$ is the angle between the direction of the force and the direction of the neutrino flux, and $\Delta(E_\nu r)$ denotes the spread of the neutrino flux energy $E_\nu$ and the location of the test masses. For larger angles, the leading $1/r$ force will be smeared out. 
Ref.~\cite{VanTilburg:2024tst} developed a different formalism to calculate the background effects (the so-called ``wake forces''), which arrived at the same results as Ref.~\cite{Ghosh:2022nzo}.

The work in Ref.~\cite{Ghosh:2022nzo} was restricted to the spin-independent part of the neutrino force, which conserves parity.
However, the neutrino force generated via Fig.~\ref{fig:Nuforce} also contains a spin-dependent part that results in both parity-conserving and parity-violating effects.
This paper focuses on the spin-dependent part of the neutrino force, particularly its parity-violating effect, in a general neutrino background. 

The amplitude in Fig.~\ref{fig:Nuforce} is given by
\begin{align}
	\label{eq:amplitude}
	{\rm i}\mathcal{A}=-\left(\frac{{\rm i} G_F}{\sqrt{2}}\right)^2 &\left[\bar{u}_1'\gamma_\mu\left(g_V^{\chi_1}+g_A^{\chi_1} \gamma_5\right)u_1\right] \int \frac{{\rm d}^4 k}{\left(2\pi\right)^4}{\rm Tr}\left[\gamma^\mu\left(1-\gamma_5\right)S_T(k+q)\gamma^\nu\left(1-\gamma_5\right)S_T(k)\right]\nonumber\\
	\times&\left[\bar{u}_2'\gamma_\nu\left(g_V^{\chi_2}+g_A^{\chi_2} \gamma_5\right)u_2\right]/\left(4m_1 m_2\right)\;,
\end{align}
where $S_T$ is a modified propagator in the background that we discuss below.
In Eq.~\eqref{eq:amplitude}, we have used $u_i\equiv u\left(p_i,s_i\right)$ to denote the wave functions, with $p_i$ and $s_i$ the momentum and spin of particle $i$, and $q=p_1'-p_1^{}=p_2^{}-p_2'$ is the momentum transfer (see Fig.~\ref{fig:Nuforce}). In the non-relativistic (NR) limit for the external particles, we have $q\approx (0,{\bf q})$ and $q^2 \approx -\rho^2$ with $\rho \equiv \left|{\bf q}\right|$.
In addition, $m_i$ is the mass of $\chi_i$ and the normalization factor of $1/(4m_1 m_2)$ should be included in the NR limit.
The neutrino force is computed by the Fourier transform of the NR limit of that diagram, 
\begin{align}
V({\bf r})=-\int \frac{{\rm d}^3 {\bf q}}{\left(2\pi\right)^3}e^{{\rm i}{\bf q}\cdot {\bf r}}{\cal A}({\bf q})\;.
\end{align}

The modified propagator in Eq.~(\ref{eq:amplitude})  is given by
\begin{eqnarray}
	\label{eq:ST}
S_T(k)=\left(\slashed{k}+m_\nu\right)\left\{\frac{\rm i}{k^2-m_\nu^2+{\rm i}\epsilon}
-2\pi\delta(k^2-m_\nu^2)
\left[\Theta(k^0)n_+\left({\bf k}\right) + \Theta(-k^0)n_- \left({\bf k}\right)\right]\right\}\,,
\end{eqnarray}
where the first term represents the vacuum propagator while the second term represents the background corrections, $\Theta$ is the Heaviside step function, and $n_{\pm}\left({\bf k}\right)$ are the distribution functions of the neutrinos and anti-neutrinos, respectively. Eq.~(\ref{eq:ST}) can be derived using the formalism of finite temperature/density field theory~\cite{Landsman:1986uw,Notzold:1987ik,Quiros:1999jp,Kapusta:2006pm,Laine:2016hma}, though the validity of Eq.~(\ref{eq:ST}) does not necessarily require a thermal distribution of the neutrino background (see Appendix A of Ref.~\cite{Ghosh:2022nzo} for a more intuitive derivation).

In the NR approximation for the external particles, it is convenient to work in the Pauli-Dirac basis where we have
\begin{align}
\label{eq:pauli-Dirac}
u_{1} = \sqrt{2 m_1}\left(
\begin{matrix}
		\xi_1\\
		\frac{\boldsymbol{\sigma}\cdot \bf{p}_{1}}{2 m_1}\xi_1
	\end{matrix}
	\right)\;,\qquad
	u_{2} = \sqrt{2 m_2}\left(
	\begin{matrix}
		\xi_2\\
		\frac{\boldsymbol{\sigma}\cdot \bf{p}_{2}}{2 m_2}\xi_2
	\end{matrix}
	\right)\;,
\end{align}
where $\boldsymbol{\sigma}\equiv (\sigma_1,\sigma_2,\sigma_3)$ are the Pauli matrices and $\xi_i$ are two-component constant spinors of $\chi_i$ (for $i=1,2$), which characterize the spins of the incident particles. For the wave functions of $u_i'$, one only needs to replace ${\bf p}_i^{}$ with ${\bf p}_i'$ in Eq.~(\ref{eq:pauli-Dirac}). Then, we obtain the wave-function contribution at the leading order of the NR approximation (i.e., up to the linear term of the velocity of external particles):
\begin{align}
	\label{eq:wavefunction}
W_{\mu\nu}\equiv \left[\bar{u}_1'\gamma_\mu\left(g_V^{\chi_1}+g_A^{\chi_1} \gamma_5\right)u_1^{}\right]\left[\bar{u}_2'\gamma_\nu\left(g_V^{\chi_2}+g_A^{\chi_2} \gamma_5\right)u_2^{}\right]/\left(4m_1 m_2\right)\equiv \Sigma_\mu^{\chi_1} \Sigma_\nu^{\chi_2}\;,
\end{align}
with
\begin{align}
	\label{eq:wavefunction-Sigma}
\Sigma_\mu^{\chi_1} &= \left(g_V^{\chi_1} + g_A^{\chi_1}\left(\boldsymbol{\sigma}_1 \cdot \boldsymbol{v}_1 + \frac{\boldsymbol{\sigma}_1 \cdot \bf{q}}{2m_1}\right), -g_A^{\chi_1} \boldsymbol{\sigma_1}-g_V^{\chi_1}\left(\boldsymbol{v}_1 + \frac{\bf{q}}{2m_1}+{\rm i}\frac{\boldsymbol{\sigma}_1\times\bf{q}}{2m_1}\right)\right)\;,\nonumber\\
\Sigma_\mu^{\chi_2} &=\left(g_V^{\chi_2}+g_A^{\chi_2}\left(\boldsymbol{\sigma}_2\cdot \boldsymbol{v}_2-\frac{\boldsymbol{\sigma}_2\cdot\bf{q}}{2m_2}\right),-g_A^{\chi_2} \boldsymbol{\sigma}_2-g_V^{\chi_2}\left(\boldsymbol{v}_2 - \frac{\bf{q}}{2m_2}-{\rm i}\frac{\boldsymbol{\sigma}_2\times\bf{q}}{2m_2}\right)\right)\;,
\end{align}
and
\begin{equation}
\boldsymbol{v}_i\equiv \frac{{\bf p}_i}{m_i}, \qquad \boldsymbol{\sigma}_i\equiv \xi_i^\dagger \boldsymbol{\sigma} \xi_i, \qquad i=1,2\;.
\end{equation}

The amplitude in Eq.~(\ref{eq:amplitude}) can be decomposed into the wave-function part $W_{\mu\nu}$ and the loop-integral factor $I^{\mu\nu}$,
\begin{align}
{\cal A}\left({\bf q}\right) = -4 G_F^2 W_{\mu\nu} I^{\mu\nu}\;,
\end{align}
with
\begin{align}
	\label{eq:loopintegral}
	I^{\mu\nu} = \frac{{\rm i}}{8} \int \frac{{\rm d}^4 k}{\left(2\pi\right)^4}{\rm Tr}\left[\gamma^\mu\left(1-\gamma_5\right)S_T(k+q)\gamma^\nu\left(1-\gamma_5\right)S_T(k)\right]\;.
\end{align}
The loop integral in Eq.~(\ref{eq:loopintegral}) can be split into the vacuum and the background contributions,
\begin{align}
I^{\mu\nu}_{}=I^{\mu\nu}_{0}+I^{\mu\nu}_{\rm bkg}\;.
\end{align}
In what follows, we shall analyze $I^{\mu\nu}_{0}$ and $I^{\mu\nu}_{\rm bkg}$ separately. Correspondingly, we decompose the amplitude and the potential into the following two parts:
\begin{align}
{\cal A}\left({\bf q}\right)={\cal A}_0\left({\bf q}\right) + {\cal A}_{\rm bkg}\left({\bf q}\right)\;,\quad
V\left({\bf r}\right) = V_0\left({\bf r}\right)+V_{\rm bkg}\left({\bf r}\right)\;,
\end{align}
where the subscripts indicate that they are proportional to $I_0^{\mu\nu}$ or $I_{\rm bkg}^{\mu\nu}$.

\subsection{The neutrino force in vacuum}

When both propagators in Eq.~(\ref{eq:loopintegral}) take the vacuum part, the integral gives the vacuum contribution to the neutrino force:
\begin{align}
	\label{eq:vacuumintegral}
I_{0}^{\mu\nu} &= -\frac{{\rm i}}{8}\int \frac{{\rm d}^4 k}{\left(2\pi\right)^4}\frac{{\rm Tr}\left[\gamma^\mu\left(1-\gamma_5\right)\left(\slashed{k}+\slashed{q}\right)\gamma^\nu\left(1-\gamma_5\right)\slashed{k}\right]}{k^2\left(k+q\right)^2}\nonumber\\
&=-\frac{1}{144\pi^2}\left[5+3\Delta_{\rm E}+3\log\left(\frac{\mu^2}{-q^2}\right)\right]\left(q^\mu q^\nu
-q^2 g^{\mu\nu}\right)\;,
\end{align}
where we have neglected the tiny neutrino mass for simplicity, 
$\mu$ is the renormalization scale and $\Delta_{\rm E}\equiv 1/\epsilon-\gamma_{\rm E}+\log\left(4\pi\right)$ with $\gamma_{\rm E}\approx 0.557$ the Euler-Mascheroni constant. (For the complete expression of the parity-violating neutrino force in vacuum, which includes finite neutrino masses and lepton flavor mixing, see Ref.~\cite{Ghosh:2019dmi}). The term proportional to $g^{00}$  in Eq.~(\ref{eq:vacuumintegral}) leads to the spin-independent potential of the neutrino force:
\begin{align}
	\label{eq:vacuumSI}
V_0^{\rm SI}(r) = g_V^{\chi_1} g_V^{\chi_2}V_0(r)\;,
\end{align}
with $V_0(r)$ given by Eq.~(\ref{eq:2nu}).
Note that the terms depending on $\mu$ and $1/\epsilon$ vanish after taking the Fourier transform since they do not have a branch cut on the complex plane of $q$. The $g^{ii}, (i = 1, 2, 3)$ and $q^{\mu}q^{\nu}$ terms in Eq.~(\ref{eq:vacuumintegral}), using the results of Fourier transform in Appendix~\ref{app:Fourier},  give rise to a spin-dependent parity-conserving (SD-PC) contribution to the potential:
\begin{align}
	\label{eq:vacuumSD}
V_0^{\text{SD-PC}}({\bf r})=\frac{1}{2}g_A^{\chi_1}g_A^{\chi_2}\left[5\left(\boldsymbol{\sigma}_1\cdot \bf{\hat{r}}\right)\left(\boldsymbol{\sigma}_2\cdot \bf{\hat{r}}\right)-3\left(\boldsymbol{\sigma}_1\cdot\boldsymbol{\sigma}_2\right)\right]V_0(r)\;,
\end{align}
with ${\bf \hat{r}}\equiv {\bf r}/r$.
Note that in Eq.~\eqref{eq:vacuumSD} $\boldsymbol{\sigma}_1$ and $\boldsymbol{\sigma}_2$ are axial vectors and  ${\bf \hat{r}}$ is a vector. Each term contains an even number of axial vectors, implying that the potential is invariant under parity transformation. 

Here, we would like to clarify the connection between spin dependence and parity violation. A force being spin-dependent does \emph{not} necessarily mean that it is parity-violating, but a parity-violating force is always spin-dependent.  The latter is labeled as SD-PV throughout this work. 

In the vacuum case, the SD-PV part of the force comes from the sub-leading term in $v$ of the contraction between $W_{\mu\nu}$ and the $g^{\mu\nu}$ term in Eq.~(\ref{eq:vacuumintegral}). 
This part 
has been calculated in Ref.~\cite{Ghosh:2019dmi},
\begin{align}
	\label{eq:V0PV}
V_0^{\text{SD-PV}} ({\bf r}) =  &\left[H_{11}\left(\boldsymbol{\sigma}_1\cdot\boldsymbol{v}_1\right)+H_{12}\left(\boldsymbol{\sigma}_1\cdot \boldsymbol{v}_2\right)+H_{21}\left(\boldsymbol{\sigma}_2\cdot \boldsymbol{v}_1\right)+H_{22}\left(\boldsymbol{\sigma}_2\cdot \boldsymbol{v}_2\right)\right.\nonumber\\
&\left.+C\left(\boldsymbol{\sigma}_1\times\boldsymbol{\sigma}_2\right)\cdot\nabla\right]V_0(r)\;,
\end{align}
where
\begin{align}
H_{11}=-H_{12}=2g_V^{\chi_2}g_A^{\chi_1}\;,\quad
H_{22}=-H_{21}=2g_V^{\chi_1}g_A^{\chi_2}\;,\quad
C=\frac{g_V^{\chi_1}g_A^{\chi_2}}{m_1}+\frac{g_V^{\chi_2}g_A^{\chi_1}}{m_2}\;.
\end{align}
Compared to Eq.~\eqref{eq:vacuumSD},  Eq.~\eqref{eq:V0PV} is suppressed by either the velocities of external particles, $\left|\boldsymbol{v}_i\right|$ (for $i=1,2$),  or by the variation of the velocities
\begin{align}
\frac{\nabla}{m_i}  \sim \frac{{\bf q}}{m_i}\sim \left| \boldsymbol{v}_i'- \boldsymbol{v}_{i}^{} \right|\;,
\end{align}
where $\boldsymbol{v}_i^{}$ and $\boldsymbol{v}_i'$ denote the incident and outgoing velocity of $\chi_i$, respectively.   Throughout this paper,  we refer to the suppression in either way as the velocity suppression.

The total neutrino force in vacuum is then obtained by adding Eqs.~(\ref{eq:vacuumSI})-(\ref{eq:V0PV}) together:
\begin{align}
V_0^{}\left({\bf r}\right) = V_0^{{\rm SI}} (r)+V_0^{\text{SD-PC}} ({\bf r})+V_0^{\text{SD-PV}} ({\bf r})\;.
\end{align}  
The first two terms are of order ${\cal O}(v^0)$, the third term is of order ${\cal O}(v)$, and we have neglected higher-ordered terms ${\cal O}(v^2)$. There are no first-order corrections to the first two terms, because any term proportional to ${\cal O}(v)$ violates parity and belongs to the third term, see Eqs.~(\ref{eq:W00PV})-(\ref{eq:WijPV}).

\subsection{The neutrino force in a general neutrino background}	
\begin{figure}[t!]
	\centering
	\includegraphics[scale=1.2]{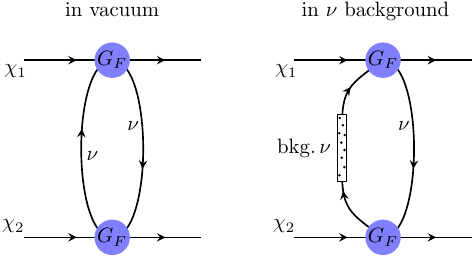}
	\caption{\label{fig:feyn}A sketch of neutrino forces in vacuum (left) and in a neutrino background (right). In vacuum, the long-range force is mediated by exchanging a pair of virtual neutrinos, both of which come from quantum fluctuations. In a background, one of the virtual neutrinos is replaced by an on-shell neutrino from the background.}
\end{figure}
We turn to the study of background corrections to the spin-dependent neutrino force. 
First, we note that there is no effect from the term when both propagators take the background part. The result is proportional to the product of two $\delta$-functions which have different arguments, and it vanishes. This can be understood as follows: If both neutrinos in Fig.~\ref{fig:feyn} are on-shell, then both of them come from the background. As a result, there is no neutrino exchanged between $\chi_1$ and $\chi_2$, and therefore, there is no force between them.

The background term comes from the cross terms in Eq.~\eqref{eq:loopintegral}, i.e., one propagator takes the vacuum part, and the other takes the background part (see the right panel of Fig.~\ref{fig:feyn}).
We find that, in general, the background integral can be decomposed into two parts,
\begin{align}
	\label{eq:IT}
	I_{\rm bkg}^{\mu\nu}\left({\bf q}\right) = I_{\rm PC}^{\mu\nu}\left({\bf q}\right) + I_{\rm PV}^{\mu\nu}\left({\bf q}\right)\;,
\end{align}
where 
\begin{align}
	\label{eq:IPC}
	I_{\rm PC}^{\mu\nu}\left({\bf q}\right) = -\int {\rm d} \widetilde{\bf k}\int {\rm d}\widetilde{k}^0\left\{\frac{2k^\mu k^\nu+k^\mu q^\nu+k^\nu q^\mu-g^{\mu\nu}\left(m_\nu^2-{\bf k}\cdot {\bf q}\right)}{2{\bf k}\cdot {\bf q}+{\bf q}^2} + \left(q\to -q\right)
	\right\}\;,
\end{align}
and
\begin{align}
	\label{eq:IPV}
	I_{\rm PV}^{\mu\nu}\left({\bf q}\right) ={\rm i}\epsilon^{\mu\nu\rho\sigma}q_\sigma \int {\rm d}\widetilde{\bf k}\int {\rm d}\widetilde{k}^0 \left[\frac{k_\rho}{2{\bf k}\cdot {\bf q}+{\bf q}^2}+ \left( q \to - q \right)\right]\;.
\end{align}
Here $\epsilon^{\mu\nu\rho\sigma}$ is the Levi-Civita tensor, and we have used the following notations:
\begin{align}
\int {\rm d}\widetilde{\bf k}&\equiv \int \frac{{\rm d}^3 {\bf k}}{\left(2\pi\right)^3}\frac{1}{2E_{\bf k}}\;,\nonumber\\
\int {\rm d}\widetilde{k}^0&\equiv \int_{-\infty}^{\infty}{\rm d}k^0\left[\delta\left(k^0-E_{\bf k}\right)\Theta\left(k^0\right)n_+ ({\bf k}) + \delta\left(k^0+E_{\bf k}\right)\Theta\left(-k^0\right)n_-({\bf k}) \right]\;,
\end{align}
with $E_{\bf k}\equiv \sqrt{{\bf k}^2+m_\nu^2}$. 
Due to the effective violation of Lorentz invariance by the neutrino background, the term proportional to $\epsilon^{\mu\nu\rho\sigma}$ is no longer vanishing as in the vacuum scenario. In fact, one can  verify that (see Appendix~\ref{app:components} for more details) 
\begin{align}
	I_{\rm PC}^{\mu\nu}\left(-{\bf q}\right)=I_{\rm PC}^{\mu\nu}\left({\bf q}\right)\;,\quad
	I_{\rm PV}^{\mu\nu}\left(-{\bf q}\right)=-I_{\rm PV}^{\mu\nu}\left({\bf q}\right)\;.
\end{align}
In vacuum, the parity-violating neutrino force only comes from the wave-function part in Eq.~(\ref{eq:wavefunction}). Technically, this is due to the fact that the vacuum loop integral in Eq.~(\ref{eq:vacuumintegral}) is invariant under the momentum reflection, i.e., $I_0^{\mu\nu}(-{\bf q})=I_0^{\mu\nu}({\bf q})$.
However, unlike in the vacuum case, the loop integral (\ref{eq:IPV}) in the background is not invariant under the momentum reflection. This leads to additional contributions to the parity-violating neutrino force, as shown below. Similarly, we can also decompose the wave functions (\ref{eq:wavefunction}) into two parts,
\begin{align}
W_{\mu\nu}^{} \left({\bf p}_i\right) = W_{\mu\nu}^{\rm PC} \left({\bf p}_i\right) + W_{\mu\nu}^{\rm PV} \left({\bf p}_i\right)\;.
\end{align}
For the specific forms of $W_{\mu\nu}^{\rm PC}$ and $W_{\mu\nu}^{\rm PV}$, we refer to Appendix~\ref{app:components}. 
Under reflection of the external momentum ${\bf p}_i\to -{\bf p}_i$, the first term is invariant while the second term changes sign
\begin{align}
	W^{\rm PC}_{\mu\nu}\left(-{\bf p}_i\right)  = W^{\rm PC}_{\mu\nu}\left({\bf p}_i\right)\;,\quad
	W^{\rm PV}_{\mu\nu}\left(-{\bf p}_i\right) = -W^{\rm PV}_{\mu\nu}\left({\bf p}_i\right)\;.
\end{align}
We note that $W^{\rm PV}_{\mu\nu}$ is suppressed by the velocities of the external particles compared with $W^{\rm PC}_{\mu\nu}$.
This is the reason the parity-violating neutrino force in vacuum, Eq.~(\ref{eq:V0PV}), is velocity suppressed, since in vacuum the only source of parity violation comes from $W_{\mu\nu}^{\rm PV}$.

The components of $I_{\rm bkg}^{\mu\nu}$ and $W_{\mu\nu}$ have been explicitly computed in Appendix~\ref{app:components}. The $(0,0)$ components of $I^{\mu\nu}_{\rm PC}$ and $W_{\mu\nu}^{\rm PC}$ give the spin-independent background amplitude,
\begin{align}
\label{eq:AbkgSI}
{\cal A}_{\rm bkg}^{\rm SI}({\bf q}) =-4G_F^2 I_{\rm PC}^{00} W_{00}^{\rm PC} = 8G_F^2 g_V^{\chi_1} g_V^{\chi_2} \int {\rm d}\widetilde{\bf k}\left[n_+ ({\bf k})+n_-({\bf k})\right]K_0\;,
\end{align}
where
\begin{align}
    \label{eq:K0}
    K_0 = \frac{\left(2{\bf k}^2+m_\nu^2\right){\bf q}^2-2\left({\bf k}\cdot{\bf q}\right)^2}{{\bf q}^4-4\left({\bf k}\cdot {\bf q}\right)^2}\;.
\end{align}
The amplitude in Eq.~(\ref{eq:AbkgSI}) is parity conserving and has been analyzed in Ref.~\cite{Ghosh:2022nzo}. 
The parity-conserving and spin-dependent background amplitude is given, to leading order in the velocity of external particles, by
\begin{align}
{\cal A}_{{\rm bkg}}^{\text{SD-PC}}({\bf q}) =&  -4G_{F}^{2}\left(I_{{\rm PC}}^{ij}W_{ij}^{{\rm PC}}+I_{{\rm PC}}^{0i}W_{0i}^{{\rm PC}}+I_{{\rm PC}}^{i0}W_{i0}^{{\rm PC}}\right)\nonumber\\
 =& +  8G_{F}^{2}g_{A}^{\chi_{1}}g_{A}^{\chi_{2}}\boldsymbol{\sigma}_{1}^{i}\boldsymbol{\sigma}_{2}^{j}\int{\rm d}\widetilde{{\bf k}}\left[n_{+}({\bf k})+n_{-}({\bf k})\right]K_{+}^{ij}\nonumber\\
&  -16G_{F}^{2}\left(g_{V}^{\chi_{1}}g_{A}^{\chi_{2}}\boldsymbol{\sigma}_{2}^{i}+g_{V}^{\chi_{2}}g_{A}^{\chi_{1}}\boldsymbol{\sigma}_{1}^{i}\right)\int{\rm d}\widetilde{{\bf k}}\left[n_{+}({\bf k})-n_{-}({\bf k})\right]K_{-}^i\;, \label{eq:AbkgSD}
\end{align}
where
\begin{align}
K_{+}^{ij} & =\frac{2k^{i}k^{j}{\bf q}^{2}-2\left(k^{i}q^{j}+k^{j}q^{i}\right)\left({\bf k}\cdot{\bf q}\right)+\delta^{ij}\left[2\left({\bf k}\cdot{\bf q}\right)^{2}+m_{\nu}^{2}{\bf q}^{2}\right]}{{\bf q}^{4}-4\left({\bf k}\cdot{\bf q}\right)^{2}}\;,\label{eq:K+}\\
K_{-}^i & =E_{{\bf k}}\frac{k^{i}{\bf q}^{2}-q^{i}\left({\bf k}\cdot{\bf q}\right)}{{\bf q}^{4}-4\left({\bf k}\cdot{\bf q}\right)^{2}}\;.\label{eq:K-}
\end{align}
Note that the $K_-$ term is suppressed by the neutrino-antineutrino asymmetry of the background, and it vanishes when the background is isotropic. 

The parity-violating background amplitude turns out to be
\begin{align}
	\label{eq:AbkgPV}
	{\cal A}_{\rm bkg}^{\text{SD-PV}}\left({\bf q}\right) = -4G_F^2\left(I_{\rm PV}^{\mu\nu}W_{\mu\nu}^{\rm PC}+I_{\rm PC}^{\mu\nu}W_{\mu\nu}^{\rm PV}\right)\;,
\end{align}
where the relevant components are given in Eqs.~(\ref{eq:I00PC})-(\ref{eq:WijPV}). 
We can split the parity-violating amplitude into two parts, according to the order of the velocity of external particles:
\begin{align}
\label{eq:AbkgPVsplit}
    {\cal A}_{\rm bkg}^{\text{SD-PV}}\left({\bf q}\right) = {\cal A}_{\rm bkg, 0}^{\text{SD-PV}}\left({\bf q}\right)+{\cal A}_{\rm bkg, 1}^{\text{SD-PV}}\left({\bf q}\right)\;.
\end{align}
The first term in Eq.~(\ref{eq:AbkgPVsplit}) does not depend on the velocity:
\begin{align}
	\label{eq:AbkgPV0}
	{\cal A}_{\rm bkg, 0}^{\text{SD-PV}}\left({\bf q}\right) =& -4G_F^2I_{\rm PV}^{\mu\nu}W_{\mu\nu}^{\rm PC}\nonumber\\
	=& +8 {\rm i}\, G_F^2 g_A^{\chi_1}g_A^{\chi_2}\epsilon^{ijk}\boldsymbol{\sigma}_1^i\boldsymbol{\sigma}_2^j \int {\rm d}\widetilde{\bf k}\left[n_+ ({\bf k})-n_-({\bf k})\right]E_{\bf k}\frac{q^k{\bf q}^2}{{\bf q}^4-4\left({\bf k}\cdot {\bf q}\right)^2}\nonumber\\
	&+8{\rm i}\,G_F^2 \left(g_V^{\chi_1}g_A^{\chi_2}\boldsymbol{\sigma}_2-g_V^{\chi_2}g_A^{\chi_1}\boldsymbol{\sigma}_1\right)^i \int {\rm d}\widetilde{\bf k}\left[n_+ ({\bf k})+n_-({\bf k})\right] \frac{ \epsilon^{ijk} k^j q^k {\bf q}^2}{{\bf q}^4-4\left({\bf k}\cdot {\bf q}\right)^2}\;.
\end{align}
We note that for neutrino-antineutrino symmetric distributions ($n_{+}=n_{-}$), the first term of Eq.~\eqref{eq:AbkgPV0} vanishes, while the second term proportional to $(n_{+}+n_{-})$ vanishes if $n_{\pm}$ are invariant under ${\bf k}\to -{\bf k}$.

The second term in Eq.~(\ref{eq:AbkgPVsplit}) is  velocity suppressed:
\begin{align}
{\cal A}_{{\rm bkg},1}^{\text{SD-PV}}\left({\bf q}\right)= & -4G_{F}^{2}I_{{\rm PC}}^{\mu\nu}W_{\mu\nu}^{{\rm PV}}\nonumber \\
= & +16G_{F}^{2}\int{\rm d}\widetilde{{\bf k}}\left[n_{+}({\bf k})+n_{-}({\bf k})\right]\left[g_{V}^{\chi_{2}}g_{A}^{\chi_{1}}\left(\boldsymbol{\sigma}_{1}^{i}\boldsymbol{v}_{1}^{i}K_{0}+\boldsymbol{\sigma}_{1}^{i}\tilde{\boldsymbol{v}}_{2}^{j}K_{+}^{ij}\right)+1\leftrightarrow2\right]\nonumber \\
& -32G_{F}^{2}\int{\rm d}\widetilde{{\bf k}}\left[n_{+}({\bf k})-n_{-}({\bf k})\right]K_{-}^{i}\left[g_{V}^{\chi_{1}}g_{V}^{\chi_{2}}\tilde{\boldsymbol{v}}_{1}^{i}+g_A^{\chi_1}g_A^{\chi_2}\left(\boldsymbol{\sigma}_1\cdot\boldsymbol{v}_1\right)\boldsymbol{\sigma}_2^i+1\leftrightarrow 2\right]\;,
\label{eq:AbkgPV1}
\end{align}
where $1\leftrightarrow 2$ denotes terms that exchange the index 1 and 2 for external particles, the reduced velocities are defined as
\begin{align}
\tilde{\boldsymbol{v}}_1\equiv  \boldsymbol{v}_1+{\rm i}\frac{\boldsymbol{\sigma}_1\times{\bf q}}{2m_1}\;,\quad  \tilde{\boldsymbol{v}}_2\equiv  \boldsymbol{v}_2-{\rm i}\frac{\boldsymbol{\sigma}_2\times{\bf q}}{2m_2}\;,
\end{align}
while $K_0$, $K_+$ and $K_-$ are given in Eqs.~(\ref{eq:K0}), (\ref{eq:K+}), and (\ref{eq:K-}), respectively. 
Note that the first term in Eq.~(\ref{eq:AbkgPV1}) is not suppressed by the neutrino-antineutrino asymmetry, while the second term vanishes for isotropic neutrino backgrounds.

The expressions in Eqs.~(\ref{eq:AbkgSI}), (\ref{eq:AbkgSD}), (\ref{eq:AbkgPV0}), and (\ref{eq:AbkgPV1}) are the comprehensive formulae we derived for the amplitude of the neutrino force in a general neutrino background. They are the main result of this work. Finally, the complete background neutrino force is given by
\begin{align}
\label{eq:V-add}
V_{\rm bkg}\left({\bf r}\right)&= V_{\rm bkg}^{\rm SI}\left(r\right)+V_{\rm bkg}^{\rm SD-PC}\left({\bf r}\right)+V_{\rm bkg,0}^{\rm SD-PV}\left({\bf r}\right)+V_{\rm bkg,1}^{\rm SD-PV}\left({\bf r}\right)\nonumber\\
&=-\int \frac{{\rm d}^3 {\bf q}}{\left(2\pi\right)^3}e^{{\rm i}{\bf q}\cdot {\bf r}}\left[{\cal A}_{\rm bkg}^{\rm SI}\left({\bf q}\right)+{\cal A}_{\rm bkg}^{\text{SD-PC}}\left({\bf q}\right)+{\cal A}_{\rm bkg,0}^{\text{SD-PV}}\left({\bf q}\right)+{\cal A}_{\rm bkg,1}^{\text{SD-PV}}\left({\bf q}\right)\right]\;,
\end{align}
where each part of the potential is determined by the Fourier transform of the corresponding amplitude.

In particular, for isotropic backgrounds $n_{\pm}({\bf k})=n_{\pm}(\kappa)$, where $\kappa\equiv \left|{\bf k}\right|$, we can first integrate the angular part and obtain the following compact expressions:
\begin{align}
    V_{\rm bkg}^{\rm SI}\left(r\right) &= -\frac{G_F^2 g_V^{\chi_1}g_V^{\chi_2}}{4\pi^3 r^5}{\cal J}_a\;,\label{eq:VSI-gen}\\
    V_{\rm bkg}^{\text{SD-PC}} \left({\bf r}\right) &= -\frac{G_F^2g_A^{\chi_1}g_A^{\chi_2}}{4\pi^3 r^5} \left[\left(\boldsymbol{\sigma}_1\cdot\boldsymbol{\sigma}_2\right){\cal J}_b+\left(\boldsymbol{\sigma}_1\cdot {\bf \hat{r}}\right)\left(\boldsymbol{\sigma}_2\cdot {\bf \hat{r}}\right){\cal J}_c\right]\;,\label{eq:VSDPC-gen}\\
    V_{\rm bkg,0}^{\text{SD-PV}} \left({\bf r}\right) &= 
{\hat{\bf r}}\cdot\left(\boldsymbol{\sigma}_1 \times \boldsymbol{\sigma}_2\right)\frac{G_F^2 g_A^{\chi_1}g_A^{\chi_2}}{4\pi^3 r^5}{\cal J}_d\;,\label{eq:VSDPV0-gen}\\
V_{\rm bkg,1}^{\text{SD-PV}} \left({\bf r}\right) &= -\frac{G_F^2g_V^{\chi_2}g_A^{\chi_1}}{2\pi^3 r^5}\left[\left(\boldsymbol{\sigma}_1\cdot\boldsymbol{v}_1\right){\cal J}_a+\left(\boldsymbol{\sigma}_1\cdot\tilde{\boldsymbol{v}}_2\right){\cal J}_b + \left(\boldsymbol{\sigma}_1\cdot {\bf \hat{r}}\right)\left(\tilde{\boldsymbol{v}}_2\cdot {\bf \hat{r}}\right){\cal J}_c\right] + 1\leftrightarrow 2 \;.\label{eq:VSDPV1-gen}
\end{align}
One should keep in mind that in the reduced velocity $\tilde{\boldsymbol{v}}_i$ in Eq.~(\ref{eq:VSDPV1-gen}), ${\bf q}$ should be replaced by~$-{\rm i}\nabla$.
The dimensionless ${\cal J}$-factors depend on the specific form of the background and can be computed as follows:
\begin{align}
    {\cal J}_a &= r \int_0^\infty {\rm d}\kappa \frac{n_+(\kappa)+n_-(\kappa)}{\sqrt{\kappa^2+m_\nu^2}}\kappa\left[\left(1+m_\nu^2 r^2\right)\sin\left(2\kappa r\right)-2\kappa r \cos\left(2\kappa r\right)\right]\;,\label{eq:Ja-gen}\\
    {\cal J}_b & = r \int_0^\infty {\rm d}\kappa \frac{n_+(\kappa)+n_-(\kappa)}{\sqrt{\kappa^2+m_\nu^2}}\kappa\left\{ 2\kappa r \cos\left(2\kappa r\right)+\left[\left(2\kappa^2+m_\nu^2\right)r^2-1\right]\sin\left(2\kappa r\right)\right\}\;,\label{eq:Jb-gen}\\
    {\cal J}_c & = 2r \int_0^\infty {\rm d}\kappa \frac{n_+(\kappa)+n_-(\kappa)}{\sqrt{\kappa^2+m_\nu^2}}\kappa\left[\left(1-\kappa^2 r^2\right)\sin\left(2\kappa r\right)-2\kappa r\cos\left(2\kappa r\right)\right]\;,\label{eq:Jc-gen}\\
    {\cal J}_d & = 2 r^2 \int_0^\infty {\rm d}\kappa \left[n_+(\kappa)-n_-(\kappa)\right]\kappa \left[\sin\left(2\kappa r\right)-\kappa r\cos\left(2\kappa r\right) \right]\;.\label{eq:Jd-gen}
\end{align}
Eqs.~(\ref{eq:VSI-gen})-(\ref{eq:Jd-gen}) are applicable to any isotropic backgrounds. To derive them, we used the Fourier transforms shown in Appendix~\ref{app:Fourier}. We provide the calculation details in Appendix~\ref{app:J-int}.

In the following two sections, we apply the general formulae derived in this section to studying several specific neutrino backgrounds: the cosmic neutrino background (C$\nu$B), the degenerate neutrino gas, and the directional neutrino beams.

\section{Isotropic backgrounds}
\label{sec:isotropic-bkg}
\subsection{Cosmic neutrino background (C$\nu$B)}\label{subsec:CvB}
The cosmic neutrino background (C$\nu$B) is ubiquitous in the universe and relatively dense compared to artificial neutrino sources.\footnote{The comparison is in terms of the number of neutrinos per volume: the C$\nu$B number density for each flavor is $56\nu/{\rm cm}^3$+$56\overline{\nu}/{\rm cm}^3$ assuming Dirac neutrinos~\cite{Workman:2022ynf}, while the reactor neutrino flux at 1 km from a 2.9 GW nuclear power plant is only $5\times 10^9/{\rm cm}^2/{\rm s}$~\cite{Kopeikin:2012zz}, corresponding to $0.17$ antineutrinos in a volume of 1 ${\rm cm}^3$. }  
The C$\nu$B is approximately isotropic.  Assuming the standard cosmological evolution, it can be approximated by the Fermi-Dirac distribution:
\begin{align}
\label{eq:FD}
n_{\pm}\left({\bf k}\right)=\frac{1}{e^{\left(\kappa\mp\mu\right)/T}+1}\;,
\end{align}
where $\mu$ and $T$ are the chemical potential and temperature of C$\nu$B, respectively. In the standard cosmology, $T\approx 1.9$ K and  $\mu/T\ll 1$.
Note that in Eq.~\eqref{eq:FD}, $T$ should not be interpreted as a temperature but as a quantity that is rescaled from the temperature of neutrino decoupling in the early universe.
Hence for massive neutrinos, Eq.~\eqref{eq:FD} is a non-thermal distribution due to $\kappa/T$ instead of $E_{{\bf k}}/T$ occurring in the exponential.      

Under certain conditions (to be discussed later), one can use the Maxwell-Boltzmann distribution to approximate the Fermi-Dirac distribution:
\begin{align}
	\label{eq:MB}
n_{\pm}\left({\bf k}\right)=\exp\left[\left(\pm \mu-\kappa\right)/T\right]\;.
\end{align}
This approximation has been frequently used in cosmological calculations for thermal species due to the simplicity of analytically computing various integrals~\cite{Gondolo:1990dk}.  And it is known that using Eq.~\eqref{eq:MB} instead of Eq.~\eqref{eq:FD} typically causes $\sim 10\%$ deviations from the true values---see e.g. Eq.~(3.6) in \cite{EscuderoAbenza:2020cmq} or Tab.~III in \cite{Luo:2020sho}. 
Therefore, in Secs.~\ref{sub:CNB-SI}-\ref{sub:CNB-SD-PV}, our results are obtained assuming the Maxwell-Boltzmann distribution. The differences caused by using the quantum statistics are discussed in Sec.~\ref{sub:CNB-FD}.

\subsubsection{The Spin-Independent (SI) part \label{sub:CNB-SI}}
Using Eq.~\eqref{eq:MB}, it is straightforward to compute the spin-independent part of the neutrino force induced by the C$\nu$B~\cite{Horowitz:1993kw,Ferrer:1998ju,Ferrer:1999ad,Ghosh:2022nzo,VanTilburg:2024tst}):
\begin{align}
\label{eq:V-CNB-SI}
    V_{\rm bkg}^{{\rm SI}}(r) &= -\frac{G_F^2 g_V^{\chi_1}g_V^{\chi_2}}{4\pi^3 r^5}{\cal J}_a^{\rm MB}\;,
\end{align}  
where
\begin{align}
{\cal J}_{a}^{\rm MB}&=2r\int_{0}^{\infty}{\rm d}\kappa\frac{\kappa e^{-\kappa/T}}{\sqrt{\kappa^{2}+m_\nu^2}}\left[\left(1+m_{\nu}^{2}r^2\right)\sin\left(2\kappa r\right)-2\kappa r\cos\left(2\kappa r\right)\right]C_{\rm L}\;.\label{eq:Ja-MB}
\end{align}
Here $C_{\rm L}=1/2$  for non-relativistic neutrinos and  $C_{\rm L}=1$ for ultra-relativistic neutrinos.
The $C_{\rm L}$ factor accounts for the effect that for non-relativistic Dirac neutrinos, half of them would be converted to the sterile state when the universe cools down to a temperature well below the neutrino mass, see Ref.~\cite{Ghosh:2022nzo} for more discussions.\footnote{We thank Ken Van Tilburg for pointing out this $1/2$ factor during discussions on Ref.~\cite{VanTilburg:2024tst}. Note that our early arXiv versions of \cite{Ghosh:2022nzo} do not contain this factor.\label{ft:CL}} 

In the relativistic ($m_{\nu}\ll T$) and non-relativistic ($m_{\nu}\gg T$) limits,   the ${\cal J}_a$  integral can be computed analytically: 
\begin{align}
    {\cal J}_a^{\rm MB} &= \frac{32 r^4 T^4}{\left(1+4r^2 T^2\right)^2}\;,\ \ \  (m_{\nu}\ll T)\;, \label{eq:Ja-MB-massless}\\
    {\cal J}_a^{\rm MB} &= \frac{4 r^4 T^3 }{m_\nu\left(1+4r^2T^2\right)^3}\left[m_\nu^2\left(1+4r^2T^2\right)+16T^2\right]\;,\ \ \  (m_{\nu}\gg T)\;. \label{eq:Ja-MB-NR}
\end{align}

\subsubsection{The Spin-Dependent Parity-Conserving (SD-PC) part  \label{sub:CNB-SD-PC}}
For the spin-dependent parity-conserving part, we only keep the first term in the amplitude~(\ref{eq:AbkgSD}) since the second term vanishes in an isotropic background. Assuming standard cosmology, the number of neutrinos and anti-neutrinos is almost the same, and thus, the chemical potential is tiny, and we neglect it. We then obtain
\begin{equation}
{\cal A}_{{\rm bkg}}^{\text{SD-PC}}\left({\bf q}\right)=8G_{F}^{2}g_{A}^{\chi_{1}}g_{A}^{\chi_{2}}\boldsymbol{\sigma}_{1}^{i}\boldsymbol{\sigma}_{2}^{j}
\int\frac{{\rm d}^{3}{\bf k}}{\left(2\pi\right)^{3}}\frac{e^{-\kappa/T}}{E_{{\bf k}}}K_{+}^{ij}\;,\label{eq:ASDK+}
\end{equation}
where $K_+$ is defined in Eq.~\eqref{eq:K+}. 

After some algebra, we obtain the corresponding neutrino force
\begin{align}
	\label{eq:VSDbkg}
V_{\rm bkg}^{\text{SD-PC}} \left({\bf r}\right) = 
-\frac{G_F^2g_A^{\chi_1}g_A^{\chi_2}}{4\pi^3 r^5}
\left[\left(\boldsymbol{\sigma}_1\cdot\boldsymbol{\sigma}_2\right){\cal J}_b^{\rm MB}+\left(\boldsymbol{\sigma}_1\cdot {\bf \hat{r}}\right)\left(\boldsymbol{\sigma}_2\cdot {\bf \hat{r}}\right){\cal J}_c^{\rm MB}\right]\;,
\end{align}
where
\begin{align}
{\cal J}_b^{\rm MB} & = 2 r \int_0^\infty {\rm d}\kappa \frac{\kappa  e^{-\kappa/T}}{\sqrt{\kappa^2+m_\nu^2}}\left\{ 2\kappa r \cos\left(2\kappa r\right)+\left[\left(2\kappa^2+m_\nu^2\right)r^2-1\right]\sin\left(2\kappa r\right)\right\}C_{\rm L}\;,\label{eq:Jb-MB}\\
{\cal J}_c^{\rm MB} & = 4r \int_0^\infty {\rm d}\kappa \frac{\kappa e^{-\kappa/T}}{\sqrt{\kappa^2+m_\nu^2}}\left[\left(1-\kappa^2 r^2\right)\sin\left(2\kappa r\right)-2\kappa r\cos\left(2\kappa r\right)\right]C_{\rm L}\;.\label{eq:Jc-MB}
\end{align}
These ${\cal J}$ integrals cannot be recast into a simple analytical form but can be computed numerically for any given values of neutrino mass $m_\nu$, temperature $T$, and distance $r$. 

In certain limits, one can obtain simple analytical results for these ${\cal J}$ integrals. 
Specifically, in the relativistic limit, these  ${\cal J}$ integrals reduce to
\begin{align}
{\cal J}_b^{\rm MB} &=16 r^4 T^4\frac{\left(1-12r^2 T^2\right)}{\left(1+4r^2 T^2\right)^3}\;,\ \ \  (m_{\nu}\ll T)\;,\label{eq:Jb-MB-massless} \\
{\cal J}_c^{\rm MB} &=16 r^4 T^4\frac{\left(1+20r^2 T^2\right)}{\left(1+4r^2 T^2\right)^3}\;,\ \ \  (m_{\nu}\ll T)\;,\label{eq:Jc-MB-massless}
\end{align}
while  in the non-relativistic limit, they are 
\begin{align}
{\cal J}_b^{\rm MB} & =\frac{4T^3 r^4}{m_\nu\left(1+4r^2 T^2\right)^4}\left[m_\nu^2\left(1+4r^2 T^2\right)^2+8T^2 \left(1-20r^2T^2\right)\right]\;, \ \ \ (m_{\nu}\gg T)\;,\label{eq:Jb-MB-NR}\\
{\cal J}_c^{\rm MB} & =\frac{32 T^5 r^4}{m_\nu}\frac{\left(1+28 r^2 T^2\right)}{\left(1+4r^2 T^2\right)^4}\;,\ \ \ (m_{\nu}\gg T)\;.\label{eq:Jc-MB-NR}
\end{align}

\subsubsection{The Spin-Dependent Parity-Violating (SD-PV) part
\label{sub:CNB-SD-PV}}
Next, we study the parity-violating part. In a general neutrino background, there are two different sources of parity violation, either coming from the loop-integral ($I_{\rm PV}^{\mu\nu}W_{\mu\nu}^{\rm PC}$) part or coming from the wave-function ($I_{\rm PC}^{\mu\nu}W_{\mu\nu}^{\rm PV}$) part, as can be seen from Eq.~(\ref{eq:AbkgPV}).  The contraction of $I_{\rm PC}^{\mu\nu}W_{\mu\nu}^{\rm PV}$ is always suppressed by the velocities of external particles or the variation of velocities. 

Let us first consider $I_{\rm PV}^{\mu\nu}W_{\mu\nu}^{\rm PC}$. 
For an isotropic background, $n_{\pm}({\bf k})=n_{\pm}({\kappa})$, the contraction of $I_{\rm PV}^{0i}W_{0i}^{\rm PC}$ and $I_{\rm PV}^{i0}W_{i0}^{\rm PC}$ vanish, as can be seen from the last line of Eq.~(\ref{eq:AbkgPV0}). The only nonzero contribution  in this contraction comes from $I_{\rm PV}^{ij}W_{ij}^{\rm PC}$, leading to the following parity-violating amplitude:
\begin{align}
	\label{eq:AbkgPV-MB}
{\cal A}_{\rm bkg,0}^{\text{SD-PV}} \left({\bf q}\right) = 8 {\rm i} G_F^2 g_A^{\chi_1}g_A^{\chi_2}\sinh\left(\frac{\mu}{T}\right)\epsilon^{ijk}\boldsymbol{\sigma}_1^i\boldsymbol{\sigma}_2^jq^k \int \frac{{\rm d}^3 {\bf k}}{\left(2\pi\right)^3}e^{-\kappa/T}\frac{{\bf q}^2}{{\bf q}^4-4\left({\bf k}\cdot {\bf q}\right)^2}\;.
\end{align}
It is interesting to notice that the parity-violating amplitude in (\ref{eq:AbkgPV-MB}) does not depend on the neutrino mass. Working out the integral and performing the Fourier transform, we obtain the parity-violating neutrino force in C$\nu$B:
\begin{align}
	\label{eq:VPVbkg0MB}
V_{\rm bkg,0}^{\text{SD-PV}} \left({\bf r}\right) = 
{\hat{\bf r}}\cdot\left(\boldsymbol{\sigma}_1 \times \boldsymbol{\sigma}_2\right)\frac{G_F^2 g_A^{\chi_1}g_A^{\chi_2}}{4\pi^3 r^5}{\cal J}_d^{\rm MB}\;,
\end{align}
where
\begin{align}
    \label{eq:Jd-MB}
    {\cal J}_d^{\rm MB} = 8r^3T^3\frac{\left(1+20r^2T^2\right)}{\left(1+4r^2T^2\right)^3}\sinh\left(\frac{\mu}{T}\right)\;.
\end{align}
This matches the result in Ref.~\cite{Horowitz:1993kw}. At short distances $r\ll 1/T$ and with small chemical potential $\mu\ll T$, Eq.~(\ref{eq:VPVbkg0MB}) scales as $V_{\rm bkg, 0}^{\text{SD-PV}}\sim G_F^2 \mu T^2/r^2$. 
Compared with the parity-violating force in vacuum (\ref{eq:V0PV}), the background effect is not suppressed by the velocity of the external particles and can also be enhanced by large temperature.

Then we consider the wave-function ($I_{\rm PC}^{\mu\nu}W_{\mu\nu}^{\rm PV}$) part, which does not require lepton asymmetry. For the isotropic background, the second term in Eq.~(\ref{eq:AbkgPV1}) vanishes. So we have
\begin{align}
\label{eq:VPVbkg1MB-int}
    {\cal A}_{\rm bkg,1}^{\text{SD-PV}} \left({\bf r}\right) = 16 G_F^2 \int \frac{{\rm d}^3{\bf k}}{\left(2\pi\right)^3}\frac{e^{-\kappa/T}}{E_k}\left[g_V^{\chi_2}g_A^{\chi_1}\left(\boldsymbol{\sigma}_1^i \boldsymbol{v}_1^iK_0 + \boldsymbol{\sigma}_1^i \tilde{\boldsymbol{v}}_2^j K_+^{ij}\right)+1\leftrightarrow 2\right]\;,
\end{align}
where $K_0$ and $K_+$ are defined in Eqs.~(\ref{eq:K0}) and (\ref{eq:K+}). After some algebra, we obtain the corresponding parity-violating neutrino force:
\begin{align}
	\label{eq:VPVbkg1MB}
V_{\rm bkg,1}^{\text{SD-PV}} \left({\bf r}\right) = -\frac{G_F^2g_V^{\chi_2}g_A^{\chi_1}}{2\pi^3 r^5}&\left[\left(\boldsymbol{\sigma}_1\cdot\boldsymbol{v}_1\right){\cal J}_a^{\rm MB}+\left(\boldsymbol{\sigma}_1\cdot\tilde{\boldsymbol{v}}_2\right){\cal J}_b^{\rm MB} + \left(\boldsymbol{\sigma}_1\cdot {\bf \hat{r}}\right)\left(\tilde{\boldsymbol{v}}_2\cdot {\bf \hat{r}}\right){\cal J}_c^{\rm MB}\right] + 1\leftrightarrow 2 \;,
\end{align}
where the complete expressions of ${\cal J}_a^{\rm MB}$, ${\cal J}_b^{\rm MB}$ and ${\cal J}_c^{\rm MB}$ are given by Eqs.~(\ref{eq:Ja-MB}), (\ref{eq:Jb-MB}) and (\ref{eq:Jc-MB}), respectively. 

\subsubsection{Maxwell-Boltzmann vs Fermi-Dirac distributions \label{sub:CNB-FD}}
The differences between the Maxwell-Boltzmann (MB) and Fermi-Dirac (FD) distributions are significant only in circumstances where the Pauli exclusion nature of fermions plays a significant role. For example, in degenerate neutrino gas, where the effect of Pauli blocking is strong, one should use the FD distribution. The degenerate limit itself is an interesting scenario, so we leave the discussions to Sec.~\ref{subsec:degenerate-bkg}. Here, we only consider the case when the degeneracy is weak (more specifically, the portion of particles below the Fermi surface, $\kappa<\mu$, makes only a very small contribution) and the chemical potential $\mu$ is small compared to all relevant quantities. We neglect the neutrino mass for simplicity in this subsection.

Under the assumptions above, we would like to address the differences between the MB and FD distributions quantitatively. It is noteworthy that the major difference can already be reflected by the number density integral $\int n_{\pm}({\bf k}) {\rm d}^3{\bf k}/(2\pi)^3$, which in the FD distribution differs by a factor of $\zeta(3)3/4\approx 0.9$ compared to the MB result. Therefore, for other integrals involved in our calculations,  differences of $\sim 10\%$  are expected. 

\begin{figure}
	\centering
	
	\includegraphics[width=0.5\linewidth]{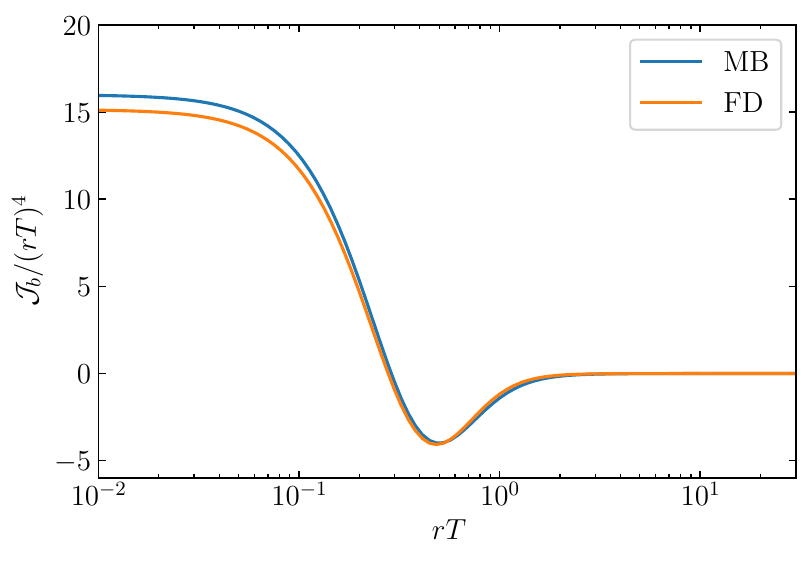}\includegraphics[width=0.5\linewidth]{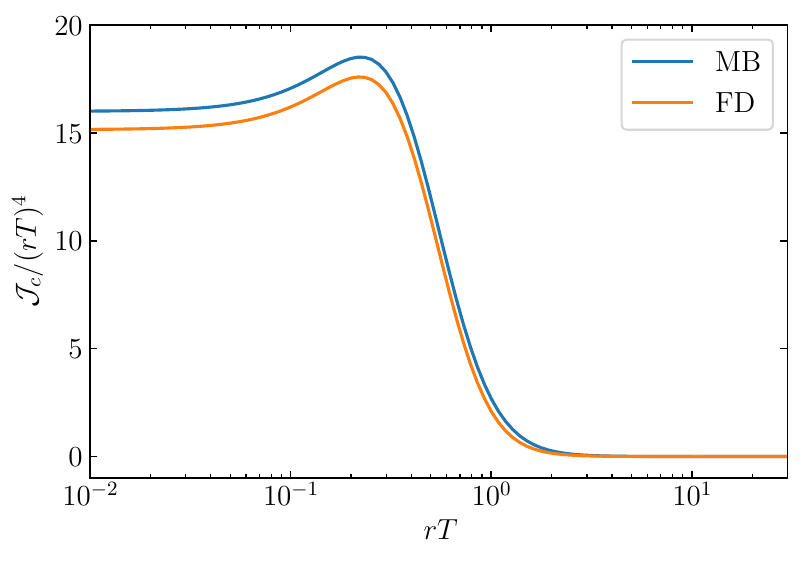}
	\caption{Comparison of the results computed using Maxwell-Boltzmann (MB) and Fermi-Dirac (FD) distributions for the ${\cal J}_b$ and ${\cal J}_c$ factors in Eq.~\eqref{eq:VSDbkg}. The MB and FD curves are produced according to Eqs.~\eqref{eq:Jb-MB-massless}-\eqref{eq:Jc-MB-massless} and \eqref{eq:Jb-FD}-\eqref{eq:Jc-FD}, respectively.
	\label{fig:MB-FD}
	 }
\end{figure}

Specifically, let us first re-calculate the parity-conserving parts using the FD distribution. 
For the SI part, the result is given by Eq.~(\ref{eq:V-CNB-SI}) with ${\cal J}_a^{\rm MB}$ being changed to
\begin{align}
    \label{eq:Ja-FD}
    {\cal J}_a^{\rm FD}  = 1-\frac{\pi r T}{\sinh\left(2\pi r T\right)} \left[1+2\pi rT\coth\left(2\pi rT\right)\right]\;.
\end{align}
For the SD-PC part, the result is given by Eq.~(\ref{eq:VSDbkg}), where ${\cal J}_b^{\rm MB}$ and ${\cal J}_c^{\rm MB}$ are replaced by
\begin{align}
{\cal J}_b^{\rm FD} & =-\frac{3}{2}+\pi rT\,\frac{-1+6\pi^{2}r^{2}T^{2}+(1+2\pi^{2}r^{2}T^{2})\cosh(4\pi rT)+2\pi rT\sinh(4\pi rT)}{2\sinh^{3}(2\pi rT)}\,,\label{eq:Jb-FD}\\
{\cal J}_c^{\rm FD} &= \frac{5}{2}-\frac{2\pi r T}{\sinh\left(2\pi r T\right)} \left[1+2\pi r T\coth\left(2\pi r T\right)\right]-\frac{\left(\pi r T\right)^3}{\sinh^3\left(2\pi r T\right)}\left[3+\cosh\left(4\pi r T\right)\right]\,.
\label{eq:Jc-FD}
\end{align}
Although the expressions for FD are much more complicated than the previous ones for MB, they are quantitatively very similar, as is shown in Fig.~\ref{fig:MB-FD}. 

We can further compute the parity-violating part with the FD distribution. 
As mentioned, here we only consider the weak degenerate limit with small $\mu$, for which we have 
\begin{align}
n_+ ({\bf k})-n_-({\bf k}) \approx \frac{2\mu}{T}\frac{ e^{\kappa/T}}{\left(1+e^{\kappa/T}\right)^2}\quad \text{($\mu\ll T$)}\;.
\end{align}
This leads to a result similar to Eq.~\eqref{eq:VPVbkg0MB} except that for FD distribution, the ${\cal J}_d$ factor is changed from Eq.~\eqref{eq:Jd-MB} to
\begin{align}
\label{eq:Jd-FD}
{\cal J}_d^{\rm FD}  = \frac{\pi r^2 T^2}{\sinh^3\left(2\pi r T\right)} \left[1+6\pi^2r^2 T^2+\left(2\pi^2r^2T^2-1\right)\cosh\left(4\pi r T\right)\right]\frac{\mu}{T}\;.
\end{align}
At short distances or low temperatures ($rT\ll 1$),  Eqs.~\eqref{eq:Jd-FD} and \eqref{eq:Jd-MB} reduce to $\frac{2\pi^2}{3} \mu T^2 r^3$ and  $8\mu  T^2 r^3$. The factor of ${\pi^2/12} \approx 0.82$  confirms our expectation that the difference between the  MB and FD cases is typical of the order of $10\%$. 

In addition, for FD distribution, we should also obtain a parity-violating term that is not suppressed by the lepton asymmetry, which is proportional to 
\begin{align}
n_+ ({\bf k})+n_-({\bf k}) \approx \frac{2}{1+e^{\kappa/T}}\quad \text{($\mu\ll T$)}\;.
\end{align}
This leads to a parity-violating force as the form of Eq.~(\ref{eq:VPVbkg1MB}) except that for FD distribution, the ${\cal J}_a$, ${\cal J}_b$ and ${\cal J}_c$ factors are given by Eqs.~(\ref{eq:Ja-FD}), (\ref{eq:Jb-FD}) and (\ref{eq:Jc-FD}), respectively.

\subsection{Degenerate neutrino gas\label{subsec:degenerate-bkg}} 
In the degenerate limit, $\mu \gg T$, the Fermi-Dirac distribution reduces to
\begin{align}
\label{eq:deglimit}
n_+ ({\bf k})\approx \left\lbrace	
\begin{aligned}
	&1\quad   ({\rm for}\; \kappa < \mu)\\
	&0\quad   ({\rm for}\; \kappa > \mu)
\end{aligned}\;,\;\;\;\qquad n_-({\bf k}) =0\;.
\right.
\end{align}
The degenerate limit is theoretically interesting and, furthermore, important to understand neutrino forces' behavior in some astrophysical environments with dense neutrino backgrounds. For instance, inside a supernovae, $T\sim {\cal O}({\rm MeV})$, while the chemical potential of neutrinos can reach $\mu\sim {\cal O}(100~{\rm MeV})$~\cite{Alford:2019kdw}. Therefore, $\mu\gg T$ could be satisfied inside the supernovae, and we can take the degenerate limit (\ref{eq:deglimit}) as a good approximation.

The degenerate limit allows relatively simple results for the neutrino forces to be obtained analytically. First of all, the spin-independent part of the potential is given by
\begin{align}
V_{{\rm bkg}}^{\text{SI}}(r)=-\frac{G_{F}^{2}g_{V}^{\chi_{1}}g_{V}^{\chi_{2}}}{4\pi^{3}r^{5}}{\cal J}_a^{\rm deg}\;,\label{eq:V-SI-deg}
\end{align}
where
\begin{align}
    {\cal J}_a^{\rm deg} = 2\sin^{2}(\mu r)-\mu r\sin(2\mu r)\;.\label{eq:Ja-deg}
\end{align}
In the short-distance limit ($\mu r\ll1$), the force behaves
as $V_{{\rm bkg}}^{\text{SI}}\sim-G_{F}^{2}\mu^{4}/r$.

The spin-dependent parity-conserving part turns out to be
\begin{align}
	\label{eq:V-SDPC-deg}
V_{\rm bkg}^{\text{SD-PC}} \left({\bf r}\right) &= 
-\frac{G_F^2g_A^{\chi_1}g_A^{\chi_2}}{4\pi^3 r^5}
\left[\left(\boldsymbol{\sigma}_1\cdot\boldsymbol{\sigma}_2\right){\cal J}_b^{\rm deg}+\left(\boldsymbol{\sigma}_1\cdot {\bf \hat{r}}\right)\left(\boldsymbol{\sigma}_2\cdot {\bf \hat{r}}\right){\cal J}_c^{\rm deg}\right]\;,
\end{align}
where
\begin{align}
{\cal J}_b^{\rm deg} &= -\frac{3}{2}+\frac{1}{2}\left(3-2\mu^2 r^2\right)\cos\left(2\mu r\right)+2\mu r\sin\left(2\mu r\right)\;,\label{eq:Jb-deg}\\
{\cal J}_c^{\rm deg} & = \frac{5}{2}+\frac{1}{2}\left(2\mu^2 r^2-5\right)\cos\left(2\mu r\right)-3\mu r \sin\left(2\mu r\right)\;.\label{eq:Jc-deg}
\end{align}
It is easy to check that at short distances ($\mu r\ll 1$), we have ${\cal J}_b^{\rm deg}={\cal J}_c^{\rm deg} = \mu^4 r^4/3$, and the force scales as $V_{\rm bkg}^{\text{SD-PC}}\sim G_F^2 \mu^4/r$.

Finally, the spin-dependent parity-violating part in this case reads:
\begin{align}
	\label{eq:V-SDPV-deg}
V_{\rm bkg}^{ \text{SD-PV}}  \left({\bf r}\right) = 
{\hat{\bf r}}\cdot\left(\boldsymbol{\sigma}_1 \times \boldsymbol{\sigma}_2\right)\frac{G_F^2 g_A^{\chi_1}g_A^{\chi_2}}{4\pi^3r^5}{\cal J}_d^{\rm deg}\;,
\end{align}
where
\begin{align}
    {\cal J}_d^{\rm deg} = \left(1-\mu^2r^2\right)\sin\left(2\mu r\right)-2\mu r\cos\left(2\mu r\right)\;.\label{eq:Jd-deg}
\end{align}
In the short-distance limit, Eq.~(\ref{eq:V-SDPV-deg}) scales as $V_{\rm bkg}^{\text{SD-PV}}\sim G_F^2 \mu^3/r^2$. 

\subsection{Summary of the results in isotropic backgrounds}
In the two subsections above, we computed the neutrino forces in isotropic backgrounds, specifically in the C$\nu$B with MB distribution and FD distribution, as well as in the degenerate neutrino gas. We found that, for any isotropic background, the neutrino forces have the same structure, as shown in Eqs.~(\ref{eq:VSI-gen})-(\ref{eq:VSDPV1-gen}). The differences are only encoded in the ${\cal J}$-factors, which are determined by the background distribution functions. In particular, the spin-independent neutrino force is only sensitive to ${\cal J}_a$, while the leading term (no velocity suppression) of the parity-violating neutrino force is only sensitive to ${\cal J}_d$. In Tab.~\ref{table:J-factors}, we summarize the results of ${\cal J}$-factors in various isotropic backgrounds that we have computed. 

\begin{table}[h]
	\centering
\begin{tabular}{ccccc}
\hline\hline
& ${\cal J}_a$ & ${\cal J}_b$& ${\cal J}_c$& ${\cal J}_d$ \\
\hline
General isotropic bkg. & \eqref{eq:Ja-gen} & \eqref{eq:Jb-gen} & \eqref{eq:Jc-gen} & \eqref{eq:Jd-gen}\\
C$\nu$B-MB & \eqref{eq:Ja-MB} & \eqref{eq:Jb-MB} & \eqref{eq:Jc-MB} & \eqref{eq:Jd-MB}\\
C$\nu$B-FD & \eqref{eq:Ja-FD} & \eqref{eq:Jb-FD} & \eqref{eq:Jc-FD} & \eqref{eq:Jd-FD}\\
Degenerate bkg. & \eqref{eq:Ja-deg} & \eqref{eq:Jb-deg} & \eqref{eq:Jc-deg} & \eqref{eq:Jd-deg}\\
\hline\hline
\end{tabular}
\caption{\label{table:J-factors}Summary of the ${\cal J}$-factors in various isotropic backgrounds. Note the results in the last two lines have neglected the neutrino mass.}
\end{table}

\section{Directional neutrino beams}
\label{sec:directional-bkg}
In this section, we consider a directional and monochromatic neutrino background, namely 
\begin{align}
	\label{eq:directional-background}
n_+\left({\bf k}\right) = \left(2\pi\right)^3\Phi_0\delta^3\left({\bf k}-{\bf k}_0\right)\;, \qquad 
n_-\left({\bf k}\right)=0\;,
\end{align}
where $\Phi_0$ is the flux of neutrinos and ${\bf k}_0$ is the momentum of neutrino flux. The energy scale of the typical neutrino beams (e.g., reactor, solar, and supernova neutrinos) is $E_\nu \sim {\cal O} ({\rm MeV})$, so one can safely neglect the neutrino mass, and thus $\left|{\bf k}_0\right|=E_\nu$.  Since the background is not isotropic, the background-induced neutrino force depends on both the distance $r$ between two external particles and the angle between ${\bf r}$ and ${\bf k}_0$ (denoted by $\alpha$).

The spin-independent neutrino force in the background of Eq.~(\ref{eq:directional-background}) has been studied in Ref.~\cite{Ghosh:2022nzo}.  The result is given by~\cite{Ghosh:2022nzo}
\begin{align}
	\label{eq:V-beam-SI}
V_{\rm bkg}^{{\rm SI}}\left(r,\alpha\right) = - \frac{G_F^2\Phi_0 E_\nu}{\pi^3}g_V^{\chi_1}g_V^{\chi_2}\,{\cal I}^{\rm SI}\left(r,\alpha\right)\;,
\end{align}
where
\begin{align}
	\label{eq:ISI-direction}
{\cal I}^{\rm SI}\left(r,\alpha\right) &\equiv \int {\rm d}^3 {\bf q}\,e^{{\rm i}{\bf q}\cdot {\bf r}}\frac{1-\xi^2}{\rho^2-4E_\nu^2\xi^2}\nonumber\\
&=\frac{\pi^2}{2r}\left(3+\cos2\alpha\right) - 2\pi E_\nu \int_{-1}^1 {\rm d}\xi\,\xi\left(1-\xi^2\right) \int_0^\pi {\rm d}\varphi \sin\left(2E_\nu r \xi \left|s_\alpha \sqrt{1-\xi^2}c_\varphi +c_\alpha \xi \right|\right)\;,
\end{align}
where $\rho\equiv \left|{\bf q}\right|$, $\xi\equiv \cos\theta$ with $\theta$ the angle between ${\bf k}_0$ and ${\bf q}$, $c_x\equiv \cos x$, $s_x \equiv \sin x$ (for $x=\alpha, \varphi$).  At long distances ($r\gg 1/E_\nu$), we have an analytical expression~\cite{Ghosh:2022nzo},
\begin{align}
    V_{\rm bkg}^{{\rm SI}}\left(r\gg E_\nu^{-1},\alpha\right) = -\frac{G_F^2\Phi_0 E_\nu}{\pi r}g_V^{\chi_1}g_V^{\chi_2}&\left\{\;\cos^2\left(\frac{\alpha}{2}\right)\cos\left[\left(1-\cos\alpha\right)E_\nu r\right]\right.\nonumber\\
    &\left.+\sin^2\left(\frac{\alpha}{2}\right)\cos\left[\left(1+\cos\alpha\right)E_\nu r\right]\right\}\;.\label{eq:V-beam-SI-long}
\end{align}
In particular, in the limit where $\alpha \to 0$ (i.e., in the direction parallel to the neutrino flux), we have $V_{\rm bkg}^{{\rm SI}}\sim G_F^2\Phi_0 E_\nu/r$, which has a significant enhancement compared with the force in vacuum, $G_F^2/r^5$, at long distances. However, such a $1/r$ force is still difficult to probe in experiments~\cite{Ghosh:2022nzo,Blas:2022ovz,VanTilburg:2024tst}, see discussions below Eq.~(\ref{eq:2nu-back}).

In the following, we calculate the spin-dependent 
parts of the neutrino force in a directional and monochromatic neutrino background. 
We first take a look at the parity-violating force.
Unlike the case of C$\nu$B, we consider the cases where the directional neutrino flux has the maximal lepton asymmetry. Thus, we keep only the leading term of the parity-violating amplitude (\ref{eq:AbkgPV0}) that is not suppressed by the velocity of external particles. Substituting Eq.~(\ref{eq:directional-background}) into Eq.~(\ref{eq:AbkgPV0}), we obtain the spin-dependent parity-violating amplitude
\begin{align}
{\cal A}_{\rm bkg,0}^{\text{SD-PV}} \left({\bf q}\right)=4{\rm i}G_F^2 \Phi_0 \epsilon^{ijk}\left[g_A^{\chi_1}g_A^{\chi_2}\boldsymbol{\sigma}_1^i\boldsymbol{\sigma}_2^j+\left(g_V^{\chi_1}g_A^{\chi_2}\boldsymbol{\sigma}_2-g_V^{\chi_2}g_A^{\chi_1}\boldsymbol{\sigma}_1\right)^i\widehat{\bf k}_0^j\right]\cdot \frac{q^k{\bf q}^2}{{\bf q}^4-4\left({\bf k}\cdot {\bf q}\right)^2}\;.
\end{align}
The corresponding parity-violating neutrino force turns out to be
\begin{align}
\label{eq:V-beam-PV-gen}
V_{\rm bkg,0}^{\text{SD-PV}}\left(r,\alpha\right)=-\frac{G_F^2\Phi_0}{2\pi^3}\left[g_A^{\chi_1}g_A^{\chi_2}\left(\boldsymbol{\sigma}_1\times\boldsymbol{\sigma}_2\right)+\left(g_V^{\chi_1}g_A^{\chi_2}\boldsymbol{\sigma}_2-g_V^{\chi_2}g_A^{\chi_1}\boldsymbol{\sigma}_1\right)\times\widehat{\bf k}_0\right]\cdot \nabla\,{\cal I}^{\rm PV}\;,
\end{align}
with $\widehat{\bf k}_0\equiv {\bf k}_0/E_\nu$, and
\begin{align}
	\label{eq:IPVdirection}
	{\cal I}^{\rm PV}\left(r,\alpha\right)& \equiv \int {\rm d}^3 {\bf q}\,e^{{\rm i}{\bf q}\cdot {\bf r}}\frac{1}{\rho^2-4E_\nu^2\xi^2}\nonumber\\
	&=\frac{2\pi^2}{r} - 2\pi E_\nu \int_{-1}^1 {\rm d}\xi\,\xi \int_0^\pi {\rm d}\varphi \sin\left(2E_\nu r \xi \left|s_\alpha \sqrt{1-\xi^2}c_\varphi +c_\alpha \xi \right|\right)\;.
\end{align}
The integral in Eq.~(\ref{eq:IPVdirection}) does not have a simple analytical expression in the most general case, but it can be computed numerically. In the following two cases, the integral can be worked out analytically.
\begin{itemize}
\item 
In the direction that is parallel to the neutrino flux, where $\alpha = 0$, we find
\begin{align}
	{\cal I}^{\rm PV}\left(r,\alpha=0\right) = \frac{\pi^2}{r}\left[1+\cos\left(2E_\nu r\right)\right]\;.
\end{align}
In this case the second term in the bracket of Eq.~(\ref{eq:V-beam-PV-gen}) vanishes, and the parity-violating neutrino force is reduced to
\begin{align}
\label{eq:V-beam-alpha0}
V_{\rm bkg,0}^{\text{SD-PV}}\left(r,\alpha=0\right) = \hat{\bf r}\cdot\left(\boldsymbol{\sigma}_1 \times \boldsymbol{\sigma}_2\right)\frac{G_F^2\Phi_0}{2\pi r^2}g_A^{\chi_1}g_A^{\chi_2} \left[1+\cos\left(2E_\nu r\right)+2E_\nu r\sin\left(2E_\nu r\right)\right]\;.
\end{align}

\item  At long distances $r\gg 1/E_\nu$, we can work out Eq.~(\ref{eq:IPVdirection}) analytically for arbitrary $\alpha$:
\begin{align}
	\label{eq:IPV-longrange}
{\cal I}^{\rm PV}\left(r\gg E_\nu^{-1},\alpha\right)=\frac{\pi^2}{r}\left\{\cos\left[\left(1-\cos\alpha\right)E_\nu r\right]+\cos\left[\left(1+\cos\alpha\right)E_\nu r\right]\right\}\;.
\end{align}
In this case, the parity-violating neutrino force is given by
\begin{align}
\label{eq:V-beam-large-r}
&V_{\rm bkg,0}^{\text{SD-PV}}\left(r\gg E_\nu^{-1},\alpha\right)=\frac{G_F^2\Phi_0}{2\pi r^2}\left[g_A^{\chi_1}g_A^{\chi_2}\left(\boldsymbol{\sigma}_1\times\boldsymbol{\sigma}_2\right)+\left(g_V^{\chi_1}g_A^{\chi_2}\boldsymbol{\sigma}_2-g_V^{\chi_2}g_A^{\chi_1}\boldsymbol{\sigma}_1\right)\times\widehat{\bf k}_0\right]\cdot \hat{{\bf r}}\nonumber\\
&\times \left\{\cos\left[\left(1-\cos\alpha\right)E_\nu r\right]+\cos\left[\left(1+\cos\alpha\right)E_\nu r\right]\right.\nonumber\\
&\left.+E_\nu r \left(1-\cos\alpha\right)\sin\left[\left(1-\cos\alpha\right)E_\nu r\right]+E_\nu r \left(1+\cos\alpha\right)\sin\left[\left(1+\cos\alpha\right)E_\nu r\right]
\right\}\;.
\end{align}
\end{itemize}
For small $\alpha$, Eq.~(\ref{eq:IPV-longrange}) is reduced to
\begin{align}
{\cal I}^{\rm PV}\left(r\gg E_\nu^{-1}, \alpha\ll 1\right) \approx \frac{\pi^2}{r}\left[\cos\left(\frac{\alpha^2 E_\nu r}{2}\right)+\cos\left(2E_\nu r\right)\right]\;.
\end{align}
We see that for $\alpha^2 \lesssim 1/\left(E_\nu r\right)$, we have
\begin{equation}
	\label{eq:VPV-parallel}
V_{\rm bkg,0}^{\text{SD-PV}}\sim \frac{G_F^2\Phi_0}{r^2}\;.
\end{equation}

For completeness, we also give the result of the next-leading order (velocity-suppressed) parity-violating force, as well as the spin-dependent parity-conserving (SD-PC) force in the directional background: 
\begin{align}
V_{\rm bkg,1}^{\text{SD-PV}}\left(r,\alpha\right) =&  -\frac{G_F^2 \Phi_0 E_\nu}{\pi^3}\left\{g_V^{\chi_2}g_A^{\chi_1}\left[2\left(\boldsymbol{\sigma}_1\cdot \boldsymbol{v}_1\right){\cal I}^{\rm SI}+\left(\boldsymbol{\sigma}_1\cdot \tilde{\boldsymbol{v}}_2\right){\cal I}_a+\left(\boldsymbol{\sigma}_1\cdot \nabla\right)\left(\tilde{\boldsymbol{v}}_2\cdot \nabla\right){\cal I}_b\right]\right.\nonumber\\
&\left.-2\left[g_V^{\chi_1}g_V^{\chi_2}\left(\tilde{\boldsymbol{v}}_1\cdot \widehat{{\bf k}}_0\right)+g_A^{\chi_1}g_A^{\chi_2}\left(\boldsymbol{\sigma}_1\cdot \boldsymbol{v}_1 \right)\left(\boldsymbol{\sigma}_2\cdot \widehat{{\bf k}}_0\right)\right]{\cal I}^{\rm PV}\right.\nonumber\\
&\left. -2\left[g_V^{\chi_1}g_V^{\chi_2}\left(\tilde{\boldsymbol{v}}_1\cdot \nabla\right)+g_A^{\chi_1}g_A^{\chi_2}\left(\boldsymbol{\sigma}_1\cdot \boldsymbol{v}_1 \right)\left(\boldsymbol{\sigma}_2\cdot \nabla\right)\right]{\cal I}_c
\right\}+ 1\leftrightarrow 2\;,
\label{eq:V-beam-PV1}
\end{align}
and
\begin{align}
    V_{\rm bkg}^{\text{SD-PC}}\left(r,\alpha\right) =- \frac{G_F^2\Phi_0 E_\nu}{2\pi^3}&\left\{g_A^{\chi_1}g_A^{\chi_2}\left[\left(\boldsymbol{\sigma}_1\cdot\boldsymbol{\sigma}_2\right){\cal I}_a+\left(\boldsymbol{\sigma}_1\cdot\nabla\right)\left(\boldsymbol{\sigma}_2\cdot\nabla\right){\cal I}_b\right]\right.\nonumber\\
    &\left. -2 g_V^{\chi_2}g_A^{\chi_1}\left[\left(\boldsymbol{\sigma}_1\cdot\widehat{{\bf k}}_0\right){\cal I}^{\rm PV}+\left(\boldsymbol{\sigma}_1\cdot \nabla\right){\cal I}_c\right] - 1\leftrightarrow 2
    \right\}\;,\label{eq:V-beam-SDPC}
\end{align}
where ${\cal I}^{\rm SI}$ and ${\cal I}^{\rm PV}$ are given by Eqs.~(\ref{eq:ISI-direction}) and (\ref{eq:IPVdirection}), and 
\begin{align}
    {\cal I}_a \left(r,\alpha\right)&\equiv  \int {\rm d}^3 {\bf q}\,e^{{\rm i}{\bf q}\cdot {\bf r}}\,\frac{1+\xi^2}{\rho^2-4E_\nu^2\xi^2}\;,\\
    {\cal I}_b \left(r,\alpha\right)&\equiv  \int {\rm d}^3 {\bf q}\,e^{{\rm i}{\bf q}\cdot {\bf r}}\,\frac{1+\xi^2}{\rho^2\left(\rho^2-4E_\nu^2\xi^2\right)}\;,\\ 
    {\cal I}_c  \left(r,\alpha\right)&\equiv  {\rm i}\int {\rm d}^3 {\bf q}\,e^{{\rm i}{\bf q}\cdot {\bf r}}\,\frac{\xi}{\rho\left(\rho^2-4E_\nu^2\xi^2\right)}\;.
\end{align}
The integrals ${\cal I}_a$, ${\cal I}_b$ and ${\cal I}_c$ turn out to be
\begin{align}
{\cal I}_a\left(r,\alpha\right)&= \frac{\pi^2}{2r}\left(5-\cos2\alpha\right)- 2\pi E_\nu \int_{-1}^1 {\rm d}\xi\,\xi \left(1+\xi^2\right) \int_0^\pi {\rm d}\varphi \sin{\cal C}\;,\label{eq:Ia}\\
{\cal I}_b\left(r,\alpha\right)&=
-\frac{\pi}{2E_\nu}\int_{-1}^1{\rm d}\xi \left(\xi+\frac{1}{\xi}\right)\int_0^\pi {\rm d}\varphi \sin{\cal C}\;,\label{eq:Ib}\\
{\cal I}_c\left(r,\alpha\right)&=
-\pi \int_{-1}^{1}{\rm d}\xi\,\xi \int_0^\pi {\rm d}\varphi \sin{\cal C}\label{eq:Ic}\;,
\end{align}
where
\begin{align}
{\cal C}=2E_\nu r \xi \left|s_\alpha \sqrt{1-\xi^2} c_\varphi +c_\alpha \xi \right|\;.
\end{align}

\section{Summary of the results and comparison with known calculations\label{sec:res}}
 
The SM neutrino force contains spin-independent (SI), spin-dependent parity-conserving (SD-PC), and parity-violating (SD-PV) parts --- all of which can be potentially influenced by neutrino backgrounds. Combining calculations of our previous work~\cite{Ghosh:2019dmi,Ghosh:2022nzo} and this work, we are able to present comprehensive results of all of them, as summarized in Tab.~\ref{tab:summary}. While a few of the results have already appeared in previous calculations, a large part of the results presented here are new.
Below, we would like to compare our results with those known results in the literature and comment on some differences we have noticed.

\begin{table}[t]
	\centering
\begin{tabular}{cccc}
	\hline\hline 
	& SI & SD-PC & SD-PV\tabularnewline
	\hline 
	Vacuum &  $^*$\eqref{eq:vacuumSI}\cite{Feinberg:1968zz}& $^*$\eqref{eq:vacuumSD}\cite{Feinberg:1968zz}  &  $^*$\eqref{eq:V0PV}\cite{Ghosh:2019dmi} \tabularnewline
	General bkg. &  $^*$\eqref{eq:AbkgSI}\cite{Ghosh:2022nzo}& \eqref{eq:AbkgSD}& \eqref{eq:AbkgPV0}+(\ref{eq:AbkgPV1})\tabularnewline
        General isotropic bkg. &
    $^*$\eqref{eq:VSI-gen}\cite{Ghosh:2022nzo}&
    \eqref{eq:VSDPC-gen}&
    \eqref{eq:VSDPV0-gen}+\eqref{eq:VSDPV1-gen}\tabularnewline
	C$\nu$B (MB) &  $^*$\eqref{eq:V-CNB-SI}\cite{Horowitz:1993kw}& $^*$\eqref{eq:VSDbkg}\cite{Ferrer:1998ju}&  $^*$\eqref{eq:VPVbkg0MB}\cite{Horowitz:1993kw}+\eqref{eq:VPVbkg1MB}\tabularnewline
	C$\nu$B (FD) &  $^*$\eqref{eq:Ja-FD}\cite{Ferrer:1999ad}& \eqref{eq:Jb-FD}+\eqref{eq:Jc-FD} & \eqref{eq:Jd-FD}\tabularnewline
	Degenerate bkg. & $^*$\eqref{eq:V-SI-deg}\cite{Horowitz:1993kw} & \eqref{eq:V-SDPC-deg} &  \eqref{eq:V-SDPV-deg}\tabularnewline
	Monochromatic beam & $^*$\eqref{eq:V-beam-SI}\cite{Ghosh:2022nzo} & \eqref{eq:V-beam-SDPC} &  \eqref{eq:V-beam-PV-gen}+\eqref{eq:V-beam-PV1}\tabularnewline
	\hline\hline 
\end{tabular}
\caption{\label{tab:summary}Summary of the results for neutrino forces in vacuum and in various backgrounds. The spin-dependent results are accurate up to order ${\cal O}(v)$ and we have neglected higher-ordered terms ${\cal O}(v^2)$, where $v$ is the velocity of external particles. Terms labeled by $*$  have been (partially) derived in the literature. After each equation, we cite the paper that derived the corresponding result for the first time.  Terms without $*$ are first derived in this work.
}	
\end{table}
The vacuum neutrino force including the SI and SD-PC parts was originally computed in Ref.~\cite{Feinberg:1968zz} and our result agrees with the original one. We note here that the vacuum SI part appearing later in \cite{Hsu:1992tg,Horowitz:1993kw} differs from the original result in Ref.~\cite{Feinberg:1968zz} by a factor of two. Here we have verified that the original result in Ref.~\cite{Feinberg:1968zz} is correct, provided proper re-interpretation of its notations in the Lagrangian.

In the presence of a background, the SI neutrino force in C$\nu$B was first calculated by Ref.~\cite{Horowitz:1993kw} and subsequently by  Ref.~\cite{Ferrer:1998ju}, both assuming the Maxwell-Boltzmann distribution. Our result in Eq.~\eqref{eq:V-CNB-SI} with the ${\cal J}_a$ factor given by Eq.~\eqref{eq:Ja-MB-massless} agrees with the former for relativistic C$\nu$B, while for non-relativistic C$\nu$B our result with the ${\cal J}_a$ factor given by Eq.~\eqref{eq:Ja-MB-NR} differs from that in Ref.~\cite{Horowitz:1993kw} by a factor of two due to the $C_{\rm L}$ factor addressed below Eq.~\eqref{eq:Ja-MB}---see also footnote~\ref{ft:CL}. Ref.~\cite{Ferrer:1998ju} uses $n_{\pm}\left({\bf k}\right)=\exp\left[\left(\pm \mu-E_{\bf k}\right)/T\right]$ which we believe is incorrect in the  non-relativistic case. 
Nevertheless, their result in the relativistic case agrees with that in Ref.~\cite{Horowitz:1993kw} and also ours. Our relativistic and non-relativistic C$\nu$B results agree with Ref.~\cite{VanTilburg:2024tst}.

For the SD-PC part of the neutrino force caused by C$\nu$B, we find that Eq.~(\ref{eq:VSDbkg}) with the ${\cal J}_b$ factor in Eq.~\eqref{eq:Jb-MB-massless} matches Eq.~(27) of Ref.~\cite{Ferrer:1998ju}. The ${\cal J}_c$ factor in Eq.~\eqref{eq:Jc-MB-massless}, however, does not match the corresponding part in  Eq.~(27) of  Ref.~\cite{Ferrer:1998ju}, with the difference being   $1+20r^2T^2$ versus $7+12r^2T^2$.

For the SD-PV part of the neutrino force caused by C$\nu$B, our result given by Eqs.~\eqref{eq:VPVbkg0MB} and \eqref{eq:Jd-MB} matches the result in Ref.~\cite{Horowitz:1993kw}. The neutrino-antineutrino symmetric contribution to SD-PV [i.e.~Eq.~\eqref{eq:VPVbkg1MB}] is absent in Ref.~\cite{Horowitz:1993kw}, hence can not be compared. 

\section{Possible experimental probes\label{sec:exp}}

The spin-dependent part of the neutrino forces may cause parity-violating effects at large distance scales. So far, the largest scale at which parity violation has been observed is the atomic scale. Atomic parity violation (APV) due to the SM weak interactions has been successfully measured for various atoms including $^{133}{\rm Cs}$, $^{205}{\rm Tl}$, $^{208} {\rm Pb}$, and $^{209}{\rm Bi}$ --- see Refs.~\cite{Workman:2022ynf,Safronova:2017xyt,Wieman:2019vik} for a review. 
Current measurements have probed the SM prediction at the sub-percent level precision. At higher energy scales (corresponding to smaller distances), many parity-violating effects had been measured, for example, in the scattering of electrons with nucleons, either elastic or deep inelastic. All of these constitute an important contingent of electroweak precision measurements. However, parity-violating scattering has weaker sensitivity for probing long-range parity-violating forces than APV~\cite{Dev:2021otb}.

In APV, the parity-violating force is connected to atomic transitions
via
\begin{equation}
\langle\Psi_{f}|V|\Psi_{i}\rangle=\int\Psi_{f}^{*}({\bf r})V({\bf r})\Psi_{i}({\bf r}){\rm d}^{3}{\bf r}\thinspace,\label{eq:APV-gen}
\end{equation}
where $\Psi_{i}$ and $\Psi_{f}$ denote the initial and final electronic wave functions, and $V$ is the potential of the parity-violating force, computed in the preceding sections of this work. For the SM $Z$-mediated parity
violation, due to the heavy mass of $Z$, the potential
can be viewed as a point-like delta function~\cite{Smirnov:2019cae}:
\begin{equation}
V_{Z}\sim G_{F}\delta^{3}({\bf r})\thinspace,\label{eq:VZ}
\end{equation}
assuming that the atomic nucleus is located at ${\bf r}=0$. 
Although the potential in Eq.~\eqref{eq:VZ} vanishes 
outside the atomic nucleus, the electron wave functions can have nonzero
values at ${\bf r}=0$. In this case, Eq.~\eqref{eq:APV-gen} with
$V=V_{Z}$ gives
\begin{equation}
\langle\Psi_{f}|V_{Z}|\Psi_{i}\rangle\sim G_{F}\Psi_{f}^{*}(0)\Psi_{i}(0)\thinspace.\label{eq:VZ-0}
\end{equation}

Eq.~\eqref{eq:VZ-0} applies to the transition between the 6\ensuremath{S}
and 7\ensuremath{S} states in Cesium, one of the most studied case
of APV. If $\Psi_{f}^{*}(0)$ or $\Psi_{i}(0)$ vanishes, which occurs
for wave functions with nonzero angular momentum (i.e. the azimuthal
quantum number $\ell \neq0$), then one should take the finite nucleus
radius $r_{N}$ into account rather than viewing it as a point-like
particle. In this case, the $Z$-mediated parity violation can be
estimated by
\begin{equation}
\langle\Psi_{f}|V_{Z}|\Psi_{i}\rangle\sim G_{F}n_{N}\int_{r<r_{N}}\Psi_{f}^{*}({\bf r})\Psi_{i}({\bf r}){\rm d}^{3}{\bf r}\thinspace, \qquad n_{N}\equiv\left(\frac{4}{3}\pi r_{N}^{3}\right)^{-1},
\label{eq:VZ-1}
\end{equation}
where $n_{N}$ can
be interpreted as the nucleus number density within the radius $r_{N}$. Note that Eq.~\eqref{eq:VZ-0} or \eqref{eq:VZ-1} only takes into account the nuclear spin-independent part, which can be added coherently among nucleons in a large nucleus and is, in general, the dominant contribution. In addition to that, there are also nuclear spin-dependent terms that become important when the spin-independent contribution is suppressed.

For the neutrino force, the potential has a finite extent due to its
long-range feature, which implies that for $\ell\neq0$ states, its parity
violation effect is non-vanishing even if the nucleus is point-like.
 Therefore, to probe neutrino forces in APV, the best avenue would
be atomic states with $\ell\neq0$ since the $Z$-mediated parity violation
in such cases is suppressed for a point-like nucleus. Unfortunately,
an order-of-magnitude estimate suggests that the parity violation
effect of the neutrino forces in APV is well below the current experimental sensitivity. The estimate for the neutrino forces in vacuum can be found in
Ref.~\cite{Ghosh:2019dmi}. Below, we present the estimate for the neutrino
forces in certain backgrounds, including the C$\nu$B, the degenerate neutrino gas, and the directional neutrino beams.

As a crude approximation aiming at obtaining the order of magnitude,
we can ignore the specific form of $\Psi_{f}^{*}({\bf r})\Psi_{i}({\bf r})$
and simply assume that they are sizable within the atomic radius $R_{0}$
(for $^{133}\text{Cs}$, $R_{0}\approx2.6\times10^{-10}$ m) and exponentially
decreases at $r>R_{0}$. Under this assumption, Eq.~\eqref{eq:APV-gen}
with $V=V_{\text{bkg}}^{\text{SD-PV}}$ can be estimated as
\begin{equation}
\langle\Psi_{f}|V_{\text{bkg}}^{\text{SD-PV}}|\Psi_{i}\rangle\sim n_{e}\int V_{\text{bkg}}^{\text{SD-PV}}({\bf r})e^{-r/R_{0}}{\rm d}^{3}{\bf r}\sim\frac{n_{e}G_{F}^{2}}{\pi^{2}}\times\begin{cases}
64R_{0}^{2}T^{4}v & T \ll R_{0}^{-1}\\
8T^{2}v & T\gg R_{0}^{-1}\\
2\mu^{3}R_{0}/3 & \mu \ll R_{0}^{-1}\\
\pi\mu^{2}/2\thinspace & \mu\gg R_{0}^{-1}
\end{cases}\ ,\label{eq:APV-nu}
\end{equation}
where $n_{e}\equiv\left(\frac{4}{3}\pi R_{0}^{3}\right)^{-1}$ is
roughly the probability density of the electron cloud. For the four
cases in Eq.~\eqref{eq:APV-nu}, the first two are obtained assuming
there is no neutrino-antineutrino asymmetry, and using Eq.~\eqref{eq:VPVbkg1MB}
in which we take  $\boldsymbol{v}_{1}\sim v$, $\tilde{\boldsymbol{v}}_{2}\sim0$, and all couplings to be ${\cal O}(1)$.
The last two are obtained using Eq.~\eqref{eq:V-SDPV-deg}. 

Comparing the first result in Eq.~\eqref{eq:APV-nu} with Eq.~\eqref{eq:VZ-0}
and taking $R_{0}\approx2.6\times10^{-10}$ m, $\Psi_{f}^{*}(0)\Psi_{i}(0)\sim n_{e}$,
and $T\sim1.9$ K, we obtain
\begin{equation}
\frac{\langle\Psi_{f}|V_{\text{bkg}}^{\text{SD-PV}}|\Psi_{i}\rangle}{\langle\Psi_{f}|V_{Z}|\Psi_{i}\rangle}\sim\frac{64G_{F}R_{0}^{2}T^{4}v}{\pi^{2}}\sim10^{-43}v \quad (\text{C$\nu$B})\;.\label{eq:V-ratio}
\end{equation}
Given that the current experiments are only able to probe $\langle\Psi_{f}|V_{Z}|\Psi_{i}\rangle$
at percent or permille level, Eq.~\eqref{eq:V-ratio} implies that
the parity-violating effect of the neutrino force is far below detectability.
Even if the atom is soaked in degenerate neutrino gas, the effect
is insignificant unless $\mu$ could exceed $\sim 23$ GeV for which
the ratio according to Eq.~\eqref{eq:APV-nu} might reach the permille level:
\begin{align}
\frac{\langle\Psi_{f}|V_{\text{bkg}}^{\text{SD-PV}}|\Psi_{i}\rangle}{\langle\Psi_{f}|V_{Z}|\Psi_{i}\rangle}\sim \frac{G_F \mu^2}{2\pi^2}\sim 10^{-3} \left(\frac{\mu}{23~{\rm GeV}}\right)^2\quad (\text{degenerate $\nu$ gas})\;.
\end{align}

On the other hand, taking the parity-violating force to be the one in a directional and monochromatic neutrino background and parallel to the flux  [cf. Eq.~(\ref{eq:VPV-parallel})], we obtain
\begin{align}
	\label{eq:V-ratio-beam}
	\frac{\langle\Psi_{f}|V_{\text{bkg}}^{\text{SD-PV}}|\Psi_{i}\rangle}{\langle\Psi_{f}|V_{Z}|\Psi_{i}\rangle}\sim 4\pi G_F \Phi_0 R_0 \sim 10^{-36} \left(\frac{\Phi_0}{10^{13}~{\rm cm}^{-2}{\rm s}^{-1}}\right)\quad (\text{directional $\nu$ beams})\;.
\end{align}
In order for the background-induced APV effect to exceed that mediated by $Z$, the flux needs to be larger than that of a typical reactor by 36 orders of magnitude.

In conclusion, the parity-violating effect of the neutrino force induced by neutrino backgrounds is too weak to be probed in existing APV experiments with foreseeable future improvements. This result, however, is not unexpected: The distance between two particles in APV experiments is of the atomic scale, which is much smaller than the average distance between two background neutrinos in typical neutrino backgrounds. For example, in the present C$\nu$B, the average distance between two background neutrinos is given by $n^{-1/3}\approx 0.26~{\rm cm}$, where $n\approx 56/{\rm cm}^3$ is the number density of the present-day cosmic neutrinos for each flavor by assuming standard cosmology~\cite{Workman:2022ynf}. This is seven orders of magnitude larger than the typical atomic length scale, described by the Bohr radius, $a_0\approx 5.3\times 10^{-9}~{\rm cm}$. As a result, the atomic system cannot `feel' the existence of the neutrino backgrounds, and the background effect is negligible compared with that of the vacuum force. On the other hand, to have a significant background force, we should look for the parity-violating effect at macroscopic scales (i.e., with a length scale larger than a centimeter). In addition, since the parity-violating force is spin-dependent, we should have a polarized macroscopic object to make the parity-violating effect coherent. The usual torsion balance experiments, which test the gravitational inverse-square law and the weak equivalence principle, are also sensitive to the SI neutrino force, while the background-induced SD-PV force cannot exceed that of the SI force (even suppressed by the velocity in some cases). So, the corresponding result cannot be better than that in our previous work~\cite{Ghosh:2022nzo}.

\section{Conclusions}
\label{sec:conclusion}
In this work, we performed a comprehensive study of the neutrino force in neutrino backgrounds, taking into account the effect of spin dependence that was neglected in many previous studies. In particular, we calculated the parity-violating effect caused by neutrino backgrounds and estimated its implications for atomic parity violation (APV) experiments.

The full expression of the neutrino force contains three parts: the spin-independent (SI) part, the spin-dependent parity-conserving (SD-PC) part, and the spin-dependent parity-violating (SD-PV) part. All of them were taken into account in this work, including background effects. In contrast to the vacuum case, neutrino backgrounds violate Lorentz invariance and lead to additional parity-violating terms that are not suppressed by the velocity of the external particles. 
The complete formulae of the neutrino forces applicable in arbitrary neutrino backgrounds were derived in Eqs.~(\ref{eq:AbkgSI}), (\ref{eq:AbkgSD}), (\ref{eq:AbkgPV0}), (\ref{eq:AbkgPV1}), and (\ref{eq:V-add}). In particular, for isotropic backgrounds, the results can be recast into a more compact form, as shown in Eqs.~(\ref{eq:VSI-gen})-(\ref{eq:Jd-gen}). We also applied our general formulae to specific backgrounds, including the C$\nu$B,  the degenerate neutrino gas, and the directional neutrino beams. The main results are summarized in Tab.~\ref{table:J-factors} and Tab.~\ref{tab:summary}.

We also estimated the parity-violating effect caused by neutrino forces in various neutrino backgrounds in APV experiments. We found that the effect is too small to be probed with the current experimental sensitivity of APV. This is because the atomic length scale is too small compared with the average distance between two background neutrinos in typical neutrino backgrounds. As a result, the atomic system is insensitive to the effects caused by background neutrinos. One possible way to probe the above effect is to look at the experiments at macroscopic length scales that are sensitive to the parity-violating process. 

Although the parity-violating effect caused by neutrino backgrounds is far from accessible in APV experiments, we hope that the comprehensive calculation presented in this paper is useful to future searches of long-range forces. For instance, the parity-violating effect caused by ultralight dark matter can be much more significant than neutrinos since it has a much larger number density than C$\nu$B and the mediator can be much lighter. This will be an interesting application of our formalism, which we leave for future work.

\section*{Acknowledgments}
The work of MG is supported in part by the US Department of Energy grant DE-SC0010102. YG is supported in part by the NSF grant  PHY-2014071. WT is supported by the NSF Grant No. PHY-2310224. BY is supported by the Samsung
Science Technology Foundation under Project Number
SSTF-BA2201-06. XJX is supported in part by the National Natural Science Foundation of China under grant No.~12141501
and also by the CAS Project for Young Scientists in Basic Research
(YSBR-099).
\begin{appendix}
\section{Fourier transform}
    \label{app:Fourier}
	In this appendix, we collect the formulae of Fourier transform that are useful in calculating the neutrino forces ($r\equiv \left|{\bf r}\right|$, $\rho\equiv \left|{\bf q}\right|$, and $a$ is an arbitrary constant independent of ${\bf q}$):
	\begin{eqnarray}
		\int\frac{{\rm d}^3 {\bf q}}{\left(2\pi\right)^3}e^{{\rm i}{\bf q}\cdot {\bf r}}\,\frac{1}{\rho^2-a^2} &=& \frac{1}{4\pi r}\cos\left(a r\right)\;,\\
		\int\frac{{\rm d}^3 {\bf q}}{\left(2\pi\right)^3}e^{{\rm i}{\bf q}\cdot {\bf r}}\,\frac{1}{\rho\left(\rho^2-a^2\right)} &=& \frac{{\rm i}}{4\pi r a} \sin\left(a r\right)\;,\\
		\int\frac{{\rm d}^3 {\bf q}}{\left(2\pi\right)^3}e^{{\rm i}{\bf q}\cdot {\bf r}}\,\frac{1}{\rho^2\left(\rho^2-a^2\right)} &=&\frac{1}{4\pi r a^2}\left[\cos\left(ar\right)-1\right]\;,\\
		\int\frac{{\rm d}^3 {\bf q}}{\left(2\pi\right)^3}e^{{\rm i}{\bf q}\cdot {\bf r}}\,\log\rho&=& -\frac{1}{4\pi r^3} \;,\label{eq:logFourier}\\
		\int\frac{{\rm d}^3 {\bf q}}{\left(2\pi\right)^3}e^{{\rm i}{\bf q}\cdot {\bf r}}\,\rho^2 \log\rho &=& -\nabla^2 \int\frac{{\rm d}^3 {\bf q}}{\left(2\pi\right)^3}e^{{\rm i}{\bf q}\cdot {\bf r}}\,\log\rho =\frac{3}{2\pi r^5}\;,\\
		\int\frac{{\rm d}^3 {\bf q}}{\left(2\pi\right)^3}e^{{\rm i}{\bf q}\cdot {\bf r}}\,q^i q^j \log\rho &=&-\partial^i \partial^j \int\frac{{\rm d}^3 {\bf q}}{\left(2\pi\right)^3}e^{{\rm i}{\bf q}\cdot {\bf r}}\,\log\rho=\frac{3}{4\pi r^5}\left(5 \frac{r^i r^j}{r^2}-\delta^{ij}\right)\;.
\end{eqnarray}

\section{Components of $I_{\rm bkg}^{\mu\nu}$ and $W_{\mu\nu}^{}$}
\label{app:components}
The amplitude in a general neutrino background is given by
\begin{align}
{\cal A}_{\rm bkg} = -4G_F^2 I_{\rm bkg}^{\mu\nu} W_{\mu\nu}^{}\;,
\end{align}
where $I_{\rm bkg}^{\mu\nu}$ [see Eqs.~(\ref{eq:IT})-(\ref{eq:IPV})] and $W_{\mu\nu}$ [see Eqs.~(\ref{eq:wavefunction})-(\ref{eq:wavefunction-Sigma})] denote the loop-integral part and the wave-function part, respectively. It is convenient to decompose $I_{\rm bkg}^{\mu\nu}$ and $W_{\mu\nu}$ into two parts:
\begin{align}
	I_{\rm bkg}^{\mu\nu} \left({\bf q}\right) = I_{\rm PC}^{\mu\nu}\left({\bf q}\right) + I_{\rm PV}^{\mu\nu} \left({\bf q}\right)\;,\quad
	W_{\mu\nu}^{} \left({\bf p}_i\right) = W_{\mu\nu}^{\rm PC} \left({\bf p}_i\right) + W_{\mu\nu}^{\rm PV} \left({\bf p}_i\right)\;,
\end{align}
where ${\bf p}_i$ is the momentum of external particles $\chi_i$ while ${\bf q}\equiv {\bf p}_1'-{\bf p}_1^{} = {\bf p}_2^{} - {\bf p}_2'$ is the momentum transfer. The subscripts PC or PV indicate that they are invariant or will change a sign under the momentum reflection:
\begin{align}
I_{\rm PC}^{\mu\nu}\left(-{\bf q}\right)&=I_{\rm PC}^{\mu\nu}\left({\bf q}\right)\;,\quad
I_{\rm PV}^{\mu\nu}\left(-{\bf q}\right)=-I_{\rm PV}^{\mu\nu}\left({\bf q}\right)\;,\nonumber\\
W^{\rm PC}_{\mu\nu}\left(-{\bf p}_i\right) & = W^{\rm PC}_{\mu\nu}\left({\bf p}_i\right)\;,\quad
W^{\rm PV}_{\mu\nu}\left(-{\bf p}_i\right)=-W^{\rm PV}_{\mu\nu}\left({\bf p}_i\right)\;.
\end{align}

In the following, we explicitly compute all the components of $I_{\rm bkg}^{\mu\nu}$ and $W_{\mu\nu}^{}$. For the loop-integral part, we obtain (where 0 denotes the time index while $i,j=1,2,3$ denote the spatial  indices)
\begin{align}
I_{\rm PC}^{00} &= -2 \int {\rm d}\widetilde{\bf k}\left[n_+ ({\bf k})+n_-({\bf k})\right]\frac{\left(2{\bf k}^2+m_\nu^2\right){\bf q}^2-2\left({\bf k}\cdot{\bf q}\right)^2}{{\bf q}^4-4\left({\bf k}\cdot {\bf q}\right)^2}\;,\label{eq:I00PC}\\
I_{\rm PC}^{0i} & = -4 \int {\rm d}\widetilde{\bf k}\left[n_+ ({\bf k})-n_-({\bf k})\right]E_{\bf k}\frac{k^i {\bf q}^2-q^i\left({\bf k}\cdot {\bf q}\right)}{{\bf q}^4-4\left({\bf k}\cdot {\bf q}\right)^2}=I_{\rm PC}^{i0}\;,\\
I_{\rm PC}^{ij} & = -2 \int {\rm d}\widetilde{\bf k}\left[n_+ ({\bf k})+n_-({\bf k})\right]\frac{2k^i k^j {\bf q}^2-2\left(k^i q^j+k^j q^i\right)\left({\bf k}\cdot {\bf q}\right)+\delta^{ij}\left[2\left({\bf k}\cdot {\bf q}\right)^2+m_\nu^2 {\bf q}^2\right]}{{\bf q}^4-4\left({\bf k}\cdot {\bf q}\right)^2}\;,\\
I_{\rm PV}^{00} & = 0\;,\\
I_{\rm PV}^{0i} &= 2{\rm i}\int {\rm d}\widetilde{\bf k}\left[n_+ ({\bf k})+n_-({\bf k})\right]\left({\bf k}\times {\bf q}\right)^i \frac{{\bf q}^2}{{\bf q}^4-4\left({\bf k}\cdot {\bf q}\right)^2}= -I_{\rm PV}^{i0}\;,\\
I_{\rm PV}^{ij} &= -2{\rm i}\epsilon^{ijk}\int {\rm d}\widetilde{\bf k}\left[n_+ ({\bf k})-n_-({\bf k})\right]E_{\bf k}\frac{q^k{\bf q}^2}{{\bf q}^4-4\left({\bf k}\cdot {\bf q}\right)^2}\;.\label{eq:IijPV}
\end{align}
In Eqs.~(\ref{eq:I00PC})-(\ref{eq:IijPV}), to make our results more general, we have not used the specific distribution of background neutrinos $n_{\pm} ({\bf k})$. For the wave-function part, we have
\begin{align}
W_{00}^{\rm PC}&=g_V^{\chi_1}g_V^{\chi_2}\;,\label{eq:W00PC}\\
W_{0i}^{\rm PC}&=-g_V^{\chi_1}g_A^{\chi_2}\boldsymbol{\sigma}_2^i\;,\\
W_{i0}^{\rm PC}&=-g_V^{\chi_2}g_A^{\chi_1}\boldsymbol{\sigma}_1^i\;,\\
W_{ij}^{\rm PC} &= g_A^{\chi_1}g_A^{\chi_2}\boldsymbol{\sigma}_1^i\boldsymbol{\sigma}_2^j\;,\\
W_{00}^{\rm PV} &=2 g_V^{\chi_2}g_A^{\chi_1}\left(\boldsymbol{\sigma}_1\cdot\boldsymbol{v}_1\right)+2 g_V^{\chi_1}g_A^{\chi_2}\left(\boldsymbol{\sigma}_2\cdot\boldsymbol{v}_2\right)\;,\label{eq:W00PV}\\
W_{0i}^{\rm PV} &=-2g_V^{\chi_1}g_V^{\chi_2}\left(\boldsymbol{v}_2-{\rm i}\frac{\boldsymbol{\sigma}_2\times{\bf q}}{2m_2}\right)^i-2g_A^{\chi_1}g_A^{\chi_2}\left(\boldsymbol{\sigma}_1\cdot\boldsymbol{v}_1\right)\boldsymbol{\sigma}_2^i\;,\label{eq:W0iPV}\\
W_{i0}^{\rm PV} &=-2g_V^{\chi_1}g_V^{\chi_2}\left(\boldsymbol{v}_1+{\rm i}\frac{\boldsymbol{\sigma}_1\times{\bf q}}{2m_1}\right)^i-2g_A^{\chi_1}g_A^{\chi_2}\left(\boldsymbol{\sigma}_2\cdot\boldsymbol{v}_2\right)\boldsymbol{\sigma}_1^i\;,\\
W_{ij}^{\rm PV} &=2g_V^{\chi_2}g_A^{\chi_1}\boldsymbol{\sigma}_1^{i}\left(\boldsymbol{v}_2-{\rm i}\frac{\boldsymbol{\sigma}_2\times{\bf q}}{2m_2}\right)^j+2g_V^{\chi_1}g_A^{\chi_2}\boldsymbol{\sigma}_2^{j}\left(\boldsymbol{v}_1+{\rm i}\frac{\boldsymbol{\sigma}_1\times{\bf q}}{2m_1}\right)^i\;.\label{eq:WijPV}
\end{align}
Note that Eqs.~(\ref{eq:W00PC})-(\ref{eq:WijPV}) are independent of the neutrino background and are only determined by the momentum and spins of external particles. In addition, the components of $W_{\mu\nu}^{\rm PV}$ are  velocity suppressed compared with those of $W_{\mu\nu}^{\rm PC}$. 

\section{The ${\cal J}$-factors \label{app:J-int}}
In this appendix, we present the details of the derivations of the ${\cal J}$-factors in Eqs.~(\ref{eq:Ja-gen})-(\ref{eq:Jd-gen}) that are applicable to any isotropic background. 

\subsection{The SI part}
For the spin-independent (SI) part, the amplitude is given by Eq.~(\ref{eq:AbkgSI})
\begin{align}
	\label{eq:Abkg-SI-app}
	{\cal A}_{\rm bkg}^{\rm SI}({\bf q})&= 8G_F^2 g_V^{\chi_1} g_V^{\chi_2} \int {\rm d}\widetilde{\bf k}\left[n_+ ({\bf k})+n_-({\bf k})\right]\frac{\left(2{\bf k}^2+m_\nu^2\right){\bf q}^2-2\left({\bf k}\cdot{\bf q}\right)^2}{{\bf q}^4-4\left({\bf k}\cdot {\bf q}\right)^2}\nonumber\\
	& = 16G_F^2 g_V^{\chi_1} g_V^{\chi_2}\int {\rm d}\widetilde{\bf k}\left[n_+ ({\bf k})+n_-({\bf k})\right]\left[\kappa^2\left(1-z^2\right)+\frac{m_\nu^2}{2}\right]\frac{1}{\rho^2-4\kappa^2z^2}\;,
\end{align}
where $\kappa \equiv \left|{\bf k}\right|$, $\rho\equiv \left|{\bf q}\right|$, and $z\equiv \cos\theta$ with $\theta$ the angle between ${\bf k}$ and ${\bf q}$. The force is the Fourier transform of the amplitude; for isotropic backgrounds, one can perform the Fourier transform before doing the loop integral over ${\bf k}$. Using the results in Appendix~\ref{app:Fourier}, we have
\begin{align}
	\int\frac{{\rm d}^3 {\bf q}}{\left(2\pi\right)^3}e^{{\rm i}{\bf q}\cdot {\bf r}}\,\frac{1}{\rho^2-4\kappa^2 z^2} = \frac{1}{4\pi r}\cos\left(2\kappa z r\right)\;.
\end{align}
Note that both $\kappa$ and $z$ are constants when integrating over ${\bf q}$.  Therefore, the SI neutrino force turns out to be
\begin{align}
	V_{\rm bkg}^{\rm SI} (r) &= -\int\frac{{\rm d}^3 {\bf q}}{\left(2\pi\right)^3}e^{{\rm i}{\bf q}\cdot {\bf r}}	{\cal A}_{\rm bkg}^{\rm SI}({\bf q})\nonumber\\
	&=-\frac{4G_F^2}{\pi r} g_V^{\chi_1} g_V^{\chi_2}  \int {\rm d}\widetilde{\bf k}\left[n_+ ({\bf k})+n_-({\bf k})\right] \left[\kappa^2\left(1-z^2\right)+\frac{m_\nu^2}{2}\right]\cos\left(2\kappa z r\right)\nonumber\\
	& = -\frac{G_F^2}{2\pi^3 r}g_V^{\chi_1}g_V^{\chi_2}\int_0^\infty {\rm d}\kappa \frac{n_+ (\kappa)+n_-(\kappa)}{\sqrt{\kappa^2+m_\nu^2}}\kappa^2\int_{-1}^1 {\rm d}z  \left[\kappa^2\left(1-z^2\right)+\frac{m_\nu^2}{2}\right]\cos\left(2\kappa z r\right)\nonumber\\
	& = -\frac{G_F^2}{4\pi^3 r^4}g_V^{\chi_1}g_V^{\chi_2}\int_0^\infty {\rm d}\kappa \frac{n_+ (\kappa)+n_-(\kappa)}{\sqrt{\kappa^2+m_\nu^2}}\kappa\left[\left(1+m_\nu^2 r^2\right)\sin\left(2\kappa r\right)-2\kappa r \cos\left(2\kappa r\right)\right]\nonumber\\
	& = -\frac{G_F^2}{4\pi^3 r^5}g_V^{\chi_1}g_V^{\chi_2}{\cal J}_a\;,\label{eq:Vbkg-SI-app}
\end{align}
where 
\begin{align}
	{\cal J}_a = r \int_0^\infty {\rm d}\kappa \frac{n_+(\kappa)+n_-(\kappa)}{\sqrt{\kappa^2+m_\nu^2}}\kappa\left[\left(1+m_\nu^2 r^2\right)\sin\left(2\kappa r\right)-2\kappa r \cos\left(2\kappa r\right)\right]\;.\label{eq:Ja-app}
\end{align}

\subsection{The SD-PC part}
Next, we consider the spin-dependent parity-conserving (SD-PC) part. For isotropic backgrounds, the term proportional to $K_-^i$ in Eq.~(\ref{eq:AbkgSD}) vanishes, so we are left with
\begin{align}
	\label{eq:Abkg-SD-PC-app}
	{\cal A}_{{\rm bkg}}^{\text{SD-PC}}({\bf q})= 8G_{F}^{2}g_{A}^{\chi_{1}}g_{A}^{\chi_{2}}\boldsymbol{\sigma}_{1}^{i}\boldsymbol{\sigma}_{2}^{j}\int{\rm d}\widetilde{{\bf k}}\left[n_{+}({\bf k})+n_{-}({\bf k})\right]K_{+}^{ij}=
	8G_F^2 g_{A}^{\chi_{1}}g_{A}^{\chi_{2}}\boldsymbol{\sigma}_{1}^{i}\boldsymbol{\sigma}_{2}^{j}\,{\cal I}^{ij}(\bf q)\;,
\end{align}
where we have defined the following tensor:
\begin{align}
	\label{eq:Iij}
	{\cal I}^{ij} ({\bf q})\equiv \int{\rm d}\widetilde{{\bf k}} \left[n_{+}({\bf k})+n_{-}({\bf k})\right]\frac{2k^{i}k^{j}{\bf q}^{2}-2\left(k^{i}q^{j}+k^{j}q^{i}\right)\left({\bf k}\cdot{\bf q}\right)+\delta^{ij}\left[2\left({\bf k}\cdot{\bf q}\right)^{2}+m_{\nu}^{2}{\bf q}^{2}\right]}{{\bf q}^{4}-4\left({\bf k}\cdot{\bf q}\right)^{2}}\;.
\end{align}
Since ${\cal I}^{ij}$ only depends on the quadratic terms of ${\bf q}$, it can be generally decomposed into
\begin{align}
	{\cal I}^{ij}({\bf q}) = \left(\rho^2\delta^{ij}-q^iq^j\right){\cal F}_{\rm T}(\rho)+q^i q^j{\cal F}_{\rm L}(\rho)\;,
\end{align}
where ${\cal F}_{\rm T}$ and ${\cal F}_{\rm L}$ are functions of $\rho \equiv \left|\bf q\right|$ that will be determined in the following. It is useful to factor out the longitudinal part of ${\cal I}^{ij}$:
\begin{align}
	\sum_{j=1}^{3} q^j {\cal I}^{ij}  = m_\nu^2 \int{\rm d}\widetilde{{\bf k}} \left[n_{+}({\bf k})+n_{-}({\bf k})\right] \frac{q^i }{\rho^2 -4\kappa^2 z^2} =\rho^2 q^i  {\cal F}_{\rm L}(\rho)\;.
\end{align}
Therefore, we find the longitudinal part of ${\cal I}^{ij}$ is proportional to $m_\nu^2$:
\begin{align}
	{\cal F}_{\rm L}(\rho) = m_\nu^2 \int{\rm d}\widetilde{{\bf k}} \left[n_{+}({\bf k})+n_{-}({\bf k})\right] \frac{1}{\rho^2\left(\rho^2-4\kappa^2z^2\right)}\;.
\end{align}
On the other hand, the transverse part of ${\cal I}^{ij}$ can be determined by taking the trace at both sides of Eq.~(\ref{eq:Iij}):
\begin{align}
	{\rm Tr}~{\cal I}\equiv \sum_{i=1}^{3} {\cal I}^{ii} &= \int{\rm d}\widetilde{{\bf k}} \left[n_{+}({\bf k})+n_{-}({\bf k})\right]\left[2\kappa^2 \left(1+z^2\right)+3m_\nu^2\right]\frac{1}{\rho^2-4\kappa^2z^2}\, , \nonumber\\
	&= 2\rho^2 {\cal F}_{\rm T}(\rho)+\rho^2 {\cal F}_{\rm L}(\rho)\;,
\end{align}
which gives 
\begin{align}
	{\cal F}_{\rm T}(\rho)  = \frac{1}{2\rho^2} {\rm Tr}~{\cal I} -\frac{1}{2}{\cal F}_{\rm L}(\rho)\;.
\end{align}
Therefore, we obtain
\begin{align}
	{\cal I}^{ij} ({\bf q})&= \frac{1}{2}\left({\rm Tr}~{\cal I}-\rho^2{\cal F}_{\rm L}\right)\delta^{ij}+\frac{1}{2\rho^2}\left(3\rho^2{\cal F}_{\rm L}-{\rm Tr}~{\cal I}\right)q^i q^j\, ,\nonumber\\
	& = \delta^{ij} \int{\rm d}\widetilde{{\bf k}} \left[n_{+}({\bf k})+n_{-}({\bf k})\right] \left[\kappa^2\left(1+z^2\right)+m_\nu^2\right]\frac{1}{\rho^2-4\kappa^2 z^2}\, ,\nonumber\\
	&- q^i q^j \int{\rm d}\widetilde{{\bf k}} \left[n_{+}({\bf k})+n_{-}({\bf k})\right] \kappa^2 \left(1+z^2\right)\frac{1}{\rho^2\left(\rho^2-4\kappa^2 z^2\right)}\;.
\end{align}
It is convenient to integrate over ${\bf q}$ before integrating over  ${\bf k}$.
Using the Fourier transform in Appendix~\ref{app:Fourier}, we have
\begin{align}
	&\int\frac{{\rm d}^3 {\bf q}}{\left(2\pi\right)^3}e^{{\rm i}{\bf q}\cdot {\bf r}}\,\frac{q^i q^j}{\rho^2\left(\rho^2-4\kappa^2 z^2\right)} \nonumber\\
	=& -\partial^i \partial^j \int\frac{{\rm d}^3 {\bf q}}{\left(2\pi\right)^3}e^{{\rm i}{\bf q}\cdot {\bf r}}\,\frac{1}{\rho^2\left(\rho^2-4\kappa^2 z^2\right)}\nonumber\\
	=&-\frac{1}{16\pi\kappa^2 z^2}\partial^i \partial^j \left\{ \frac{1}{r}\left[\cos\left(2\kappa r z\right)-1\right]\right\}\nonumber\\
	=&-\frac{1}{16\pi\kappa^2 z^2 r^3}\left\{\hat{r}^i\hat{r}^j\left[\left(3-4\kappa^2z^2 r^2\right)\cos\left(2\kappa z r\right)+6\kappa z r \sin\left(2\kappa z r\right)-3\right]\right.\nonumber\\
	&\left. \qquad \qquad \qquad\; -\delta^{ij}\left[\cos\left(2\kappa z r\right)+2\kappa z r\sin\left(2\kappa z r\right)-1\right] \right\}\;,
\end{align}
which leads to
\begin{align}
	&\int\frac{{\rm d}^3 {\bf q}}{\left(2\pi\right)^3}e^{{\rm i}{\bf q}\cdot {\bf r}}\, {\cal I}^{ij}({\bf q}) = \frac{1}{4\pi r^3} \int{\rm d}\widetilde{{\bf k}} \left[n_{+}({\bf k})+n_{-}({\bf k})\right]\left({\cal A}\,\delta^{ij}+{\cal B}\,\hat{r}^i \hat{r}^j \right)\;,
\end{align}
where $\hat{r}^i\equiv r^i/r$ and
\begin{align}
	{\cal A} &= r^2\left[\kappa^2\left(1+z^2\right)+m_\nu^2\right]\cos\left(2\kappa z r\right)-\frac{1+z^2}{4z^2}\left[\cos\left(2\kappa z r\right)+2\kappa z r\sin\left(2\kappa z r\right)-1\right]\;,\\
	{\cal B}&=\frac{1+z^2}{4z^2}\left[\left(3-4\kappa^2z^2 r^2\right)\cos\left(2\kappa z r\right)+6\kappa z r \sin\left(2\kappa z r\right)-3\right]\;.
\end{align}
Therefore, the SD-PC part of the neutrino force is given by
\begin{align}
	\label{eq:VSDapp}
	V_{\rm bkg}^{\text{SD-PC}}({\bf r})&= -\int\frac{{\rm d}^3 {\bf q}}{\left(2\pi\right)^3}e^{{\rm i}{\bf q}\cdot {\bf r}}	{\cal A}_{\rm bkg}^{\rm SD-PC}({\bf q})\nonumber\\
	& = -8G_F^2 g_{A}^{\chi_{1}}g_{A}^{\chi_{2}}\boldsymbol{\sigma}_{1}^{i}\boldsymbol{\sigma}_{2}^{j} \int\frac{{\rm d}^3 {\bf q}}{\left(2\pi\right)^3}e^{{\rm i}{\bf q}\cdot {\bf r}}\,{\cal I}^{ij}({\bf q})\nonumber\\
	& = -\frac{G_F^2}{4\pi^3  r^3}g_A^{\chi_1}g_{A}^{\chi_{2}}\boldsymbol{\sigma}_{1}^{i}\boldsymbol{\sigma}_{2}^{j}\int_0^\infty {\rm d}\kappa \frac{n_+(\kappa)+n_-(\kappa)}{\sqrt{\kappa^2+m_\nu^2}}\kappa^2 \int_{-1}^1 {\rm d}z \left({\cal A}\,\delta^{ij}+{\cal B}\,\hat{r}^i \hat{r}^j \right)\;.
\end{align}
The integrals of ${\cal A}$ and ${\cal B}$ give
\begin{align}
	\int_{-1}^{1}{\rm d}z\,{\cal A} &= 2\cos\left(2\kappa r\right) +\frac{1}{\kappa r}\left[\left(2\kappa^2+m_\nu^2\right)r^2-1\right]\sin\left(2\kappa r\right)\;,\\
	\int_{-1}^{1}{\rm d}z\,{\cal B} &= \frac{2}{\kappa r}\left[\sin\left(2\kappa r\right)-2\kappa r\cos\left(2\kappa r\right)-\kappa^2 r^2 \sin\left(2\kappa r\right)\right]\;.
\end{align}
Substituing them back to Eq.~(\ref{eq:VSDapp}), one obtains
\begin{align}
	V_{\rm bkg}^{\text{SD-PC}} \left({\bf r}\right) = -\frac{G_F^2g_A^{\chi_1}g_A^{\chi_2}}{4\pi^3 r^5} \left[\left(\boldsymbol{\sigma}_1\cdot\boldsymbol{\sigma}_2\right){\cal J}_b+\left(\boldsymbol{\sigma}_1\cdot {\bf \hat{r}}\right)\left(\boldsymbol{\sigma}_2\cdot {\bf \hat{r}}\right){\cal J}_c\right]\;,\label{eq:Vbkg-SD-PC-app}
\end{align}
where
\begin{align}
	{\cal J}_b & = r \int_0^\infty {\rm d}\kappa \frac{n_+(\kappa)+n_-(\kappa)}{\sqrt{\kappa^2+m_\nu^2}}\kappa\left\{ 2\kappa r \cos\left(2\kappa r\right)+\left[\left(2\kappa^2+m_\nu^2\right)r^2-1\right]\sin\left(2\kappa r\right)\right\}\;,\label{eq:Jb-app}\\
	{\cal J}_c & = 2r \int_0^\infty {\rm d}\kappa \frac{n_+(\kappa)+n_-(\kappa)}{\sqrt{\kappa^2+m_\nu^2}}\kappa\left[\left(1-\kappa^2 r^2\right)\sin\left(2\kappa r\right)-2\kappa r\cos\left(2\kappa r\right)\right]\;.\label{eq:Jc-app}
\end{align}

\subsection{The SD-PV part}
The spin-dependent parity-violating (SD-PV) amplitude includes two terms, as shown in Eqs.~(\ref{eq:AbkgPV0}) and (\ref{eq:AbkgPV1}).  

We first consider the leading term of  ${\cal A}_{\rm bkg, 0}^{\text{SD-PV}}$ in Eq.~(\ref{eq:AbkgPV0}),  which is not supressed by velocity. For isotropic backgrounds, the last line of Eq.~(\ref{eq:AbkgPV0}) vanishes, so we have
\begin{align}
	{\cal A}_{\rm bkg, 0}^{\text{SD-PV}}\left({\bf q}\right) 
	&=8 {\rm i} G_F^2 g_A^{\chi_1}g_A^{\chi_2}\epsilon^{ijk}\boldsymbol{\sigma}_1^i\boldsymbol{\sigma}_2^j \int {\rm d}\widetilde{\bf k}\left[n_+ ({\bf k})-n_-({\bf k})\right]E_{\bf k}\frac{q^k{\bf q}^2}{{\bf q}^4-4\left({\bf k}\cdot {\bf q}\right)^2}\, ,\nonumber\\
	& = 4{\rm i}G_F^2 g_A^{\chi_1}g_A^{\chi_2}\epsilon^{ijk}\boldsymbol{\sigma}_1^i\boldsymbol{\sigma}_2^j \int \frac{{\rm d}^3 {\bf k}}{\left(2\pi\right)^3}\left[n_+ ({\bf k})-n_-({\bf k})\right]\frac{q^k}{\rho^2-4\kappa^2 z^2}\;.
\end{align}
Perform the Fourier transform first:
\begin{align}
	\int\frac{{\rm d}^3 {\bf q}}{\left(2\pi\right)^3}e^{{\rm i}{\bf q}\cdot {\bf r}} \frac{q^k}{\rho^2-4\kappa^2 z^2}&=-{\rm i}\partial^k \int\frac{{\rm d}^3 {\bf q}}{\left(2\pi\right)^3}e^{{\rm i}{\bf q}\cdot {\bf r}} \frac{1}{\rho^2-4\kappa^2 z^2}\, ,\nonumber\\
	& = -\frac{{\rm i}}{4\pi} \partial^k \left[\cos\left(2\kappa z r\right)\right]\, ,\nonumber\\
	& = \frac{{\rm i}}{4\pi r^2}\hat{r}^k\left[\cos\left(2\kappa z r\right)+2\kappa z r \sin\left(2\kappa z r\right)\right]\;.
\end{align}
The corresponding parity-violating force is then given by
\begin{align}
	V_{\rm bkg,0}^{\text{SD-PV}}({\bf r})&= -\int\frac{{\rm d}^3 {\bf q}}{\left(2\pi\right)^3}e^{{\rm i}{\bf q}\cdot {\bf r}}	{\cal A}_{\rm bkg,0}^{\rm SD-PV}({\bf q})\, \nonumber\\
	& \hspace*{-15mm}=  \frac{G_F^2g_A^{\chi_1}g_A^{\chi_2}}{4\pi^3 r^2}\epsilon^{ijk}\boldsymbol{\sigma}_1^i\boldsymbol{\sigma}_2^j\, \hat{r}^k_{}\int_0^\infty {\rm d}\kappa \left[n_+(\kappa)-n_-(\kappa)\right]\kappa^2 \int_{-1}^{1}{\rm d}z \left[\cos\left(2\kappa z r\right)+2\kappa z r \sin\left(2\kappa z r\right)\right]\, \nonumber\\
	& \hspace*{-15mm} = \frac{G_F^2g_A^{\chi_1}g_A^{\chi_2}}{4\pi^3 r^2}\epsilon^{ijk}\boldsymbol{\sigma}_1^i\boldsymbol{\sigma}_2^j\, \hat{r}^k_{}\int_0^\infty {\rm d}\kappa \left[n_+(\kappa)-n_-(\kappa)\right]\frac{2\kappa}{r}\left[\sin\left(2\kappa r\right)-\kappa r \cos\left(2\kappa r\right)\right]\, \nonumber\\
	& \hspace*{-15mm}= {\hat{\bf r}}\cdot\left(\boldsymbol{\sigma}_1 \times \boldsymbol{\sigma}_2\right)\frac{G_F^2g_A^{\chi_1}g_A^{\chi_2}}{4\pi^3 r^5}{\cal J}_d\;,
\end{align}
where
\begin{align}
	{\cal J}_d = 2r^2 \int_0^\infty {\rm d}\kappa \left[n_+(\kappa)-n_-(\kappa)\right]\kappa \left[\sin\left(2\kappa r\right)-\kappa r\cos\left(2\kappa r\right)\right]\;.
\end{align}
It should be emphasized that if there is no lepton asymmetry ($n_+ = n_-$), ${\cal J}_d$ will exactly vanish while ${\cal J}_a$, ${\cal J}_b$ and ${\cal J}_c$ are nonvanishing.

Then, we consider the next-leading term of the amplitude ${\cal A}_{\rm bkg, 1}^{\text{SD-PV}}$ in Eq.~(\ref{eq:AbkgPV1}), which is velocity suppressed.  For isotropic backgrounds, the term proportional to $K_-^i$ in Eq.~(\ref{eq:AbkgPV1}) vanishes, so the amplitude is reduced to
\begin{align}
	\label{eq:Abkg-SD-PV-app}
	{\cal A}_{{\rm bkg},1}^{\text{SD-PV}}\left({\bf q}\right)=  16G_{F}^{2}\int{\rm d}\widetilde{{\bf k}}\left[n_{+}({\bf k})+n_{-}({\bf k})\right]\left[g_{V}^{\chi_{2}}g_{A}^{\chi_{1}}\left(\boldsymbol{\sigma}_{1}^{i}\boldsymbol{v}_{1}^{i}K_{0}+\boldsymbol{\sigma}_{1}^{i}\tilde{\boldsymbol{v}}_{2}^{j}K_{+}^{ij}\right)+1\leftrightarrow2\right]\;.
\end{align}
Comparing Eq.~(\ref{eq:Abkg-SD-PV-app}) with Eqs.~(\ref{eq:Abkg-SI-app}) and (\ref{eq:Abkg-SD-PC-app}), we find the dependence of ${\cal A}_{\rm bkg, 1}^{\text{SD-PV}}$ on the loop menemtum ${\bf k}$ is the same as those of ${\cal A}_{\rm bkg}^{{\rm SI}}$ and ${\cal A}_{\rm bkg}^{\text{SD-PC}}$. Therefore, according to the above results in Eqs.~(\ref{eq:Vbkg-SI-app}) and (\ref{eq:Vbkg-SD-PC-app}), one can directly write down the corresponding parity-violating neutrino force
\begin{align}
	V_{\rm bkg,1}^{\text{SD-PV}} \left({\bf r}\right) = -\frac{G_F^2g_V^{\chi_2}g_A^{\chi_1}}{2\pi^3 r^5}\left[\left(\boldsymbol{\sigma}_1\cdot\boldsymbol{v}_1\right){\cal J}_a+\left(\boldsymbol{\sigma}_1\cdot\tilde{\boldsymbol{v}}_2\right){\cal J}_b + \left(\boldsymbol{\sigma}_1\cdot {\bf \hat{r}}\right)\left(\tilde{\boldsymbol{v}}_2\cdot {\bf \hat{r}}\right){\cal J}_c\right] + 1\leftrightarrow 2\;,
\end{align}
where ${\cal J}_a$, ${\cal J}_b$ and ${\cal J}_c$ are given by Eqs.~(\ref{eq:Ja-app}), (\ref{eq:Jb-app}) and (\ref{eq:Jc-app}), respectively.

\end{appendix}
	\bibliographystyle{JHEP}
	\bibliography{ref}
\end{document}